\documentclass[sigconf,nonacm]{acmart}
\AtBeginDocument{%
  }

\setcopyright{acmlicensed}
\copyrightyear{2018}
\acmYear{2018}
\acmDOI{XXXXXXX.XXXXXXX}
\acmConference[Conference acronym 'XX]{Make sure to enter the correct
  conference title from your rights confirmation email}{June 03--05,
  2018}{Woodstock, NY}
  
\acmISBN{978-1-4503-XXXX-X/2018/06}

\usepackage{bm} 
\usepackage{multirow}
\usepackage{algorithm}
\usepackage{algpseudocode}
\usepackage{pifont}
\usepackage{tcolorbox}
\tcbuselibrary{breakable, skins, listings}

\begin{document}

\def\x{{\mathbf x}}
\def\L{{\cal L}}
\def\Dcon{$\mathcal{D}_{\text{con}}$}
\def\Rmem{$\mathcal{R}_{\text{mem}}$}
\def\Aass{$\mathcal{A}_{\text{ass}}$}
\def\Agen{$\mathcal{A}_{\text{gen}}$}
\def\Arep{$\mathcal{A}_{\text{rep}}$}
\def\Amem{$\mathcal{A}_{\text{mem}}$}

\title{GRIDAI: Generating and Repairing Intrusion Detection Rules via Collaboration among Multiple LLM-based Agents}

\author{Jiarui Li}
\affiliation{
  \institution{Harbin Institute of Technology, Shenzhen}
  \city{Shenzhen}
  \state{Guangdong}
  \country{China}}
\email{24S151130@stu.hit.edu.cn}

\author{Yuhan Chai}
\affiliation{
  \institution{Guangzhou University}
  \city{Guangzhou}
  \state{Guangdong}
  \country{China}}
\email{chaiyuhan@gzhu.edu.cn}

\author{Lei Du}
\affiliation{
  \institution{Pengcheng Laboratory}
  \city{Shenzhen}
  \state{Guangdong}
  \country{China}}
\email{dul@pcl.ac.cn}

\author{Chenyun Duan}
\affiliation{
  \institution{Harbin Institute of Technology, Shenzhen}
  \city{Shenzhen}
  \state{Guangdong}
  \country{China}}
\email{200110315@stu.hit.edu.cn}

\author{Hao Yan}
\authornote{Also with Pengcheng Laboratory.}
\affiliation{
  \institution{Harbin Institute of Technology, Shenzhen}
  \city{Shenzhen}
  \state{Guangdong}
  \country{China}}
\email{yanhao@stu.hit.edu.cn}

\author{Zhaoquan Gu}
\authornote{Corresponding author and also with Pengcheng Laboratory.}
\affiliation{
  \institution{Harbin Institute of Technology, Shenzhen}
  \city{Shenzhen}
  \state{Guangdong}
  \country{China}}
\email{guzhaoquan@hit.edu.cn}

\renewcommand{\shortauthors}{Li et al.}

\begin{abstract}
Rule-based network intrusion detection systems play a crucial role in the real-time detection of Web attacks.
However, most existing works primarily focus on automatically generating detection rules for new attacks, often overlooking the relationships between new attacks and existing rules, which leads to significant redundancy within the ever-expanding ruleset.
To address this issue, we propose \textbf{GRIDAI}, a novel end-to-end framework for the automated \textbf{G}eneration and \textbf{R}epair of \textbf{I}ntrusion \textbf{D}etection rules through collaboration among multiple LLM-based agents.
Unlike traditional methods, GRIDAI first assesses the nature of incoming attack samples. 
If the sample represents a new attack type, it is used to generate a new rule. 
Otherwise, the sample is identified as a variant of an attack already covered by an existing rule and used to repair the rule by updating the corresponding signature, thereby enhancing its generalization capability.
Additionally, to mitigate syntactic and semantic errors in rules caused by LLM hallucinations, we incorporate a tool-based real-time validation mechanism and a representative attack sample maintained for each rule, enabling fully automated rule generation and repair.
Comprehensive experiments were conducted on a public dataset containing seven types of attacks and a private dataset with 43 attack types. 
The results demonstrate that GRIDAI accurately identifies the relationships between new attack samples and existing rules, efficiently generates and repairs rules to handle new attacks and variants, and effectively mitigates the impact of LLM hallucinations.

\end{abstract}



\maketitle

\section{Introduction}\label{sec:intro}
In recent years, cybersecurity threats on the Web have become increasingly frequent, causing substantial economic damage worldwide.
According to the latest annual report from the Federal Bureau of Investigation, approximately 859,532 internet crime complaints were filed in 2024, resulting in economic damages exceeding  \$16 billion\cite{fbi01}. 
Beyond direct financial losses, these cyberattacks also severely undermine the availability and reliability of Web environments, leading to data breaches and eroding user trust. 
Consequently, it is crucial to implement high-precision, real-time attack detection in Web scenarios for safeguarding the security of the online ecosystem, ensuring service stability, and protecting user activities.

Numerous solutions have been proposed for attack detection, including Network Intrusion Detection Systems, Host-based Intrusion Detection Systems, and Firewalls.
Among these, Rule-based Network Intrusion Detection Systems (RNIDS) play a crucial role due to their high detection accuracy, low false-positive rates, and ease of deployment\cite{sommer2010outside, tavallaee2010toward, vermeer2023alert}. 
This paradigm, also known as misuse detection, identifies known attack behaviors by matching real-time network traffic with predefined attack signatures, with typical examples including Suricata\footnote{https://suricata.io/} and Snort\footnote{https://www.snort.org/}.

The protective capability of RNIDS fundamentally depends on the quality of its ruleset, which is typically handcrafted by experienced security experts based on identified attack patterns in network traffic.
However, in the rapidly evolving network environment, this manual rule-design approach is becoming increasingly insufficient, often leading to significant delays in responding to emerging threats \cite{diaz2022detection}.
For instance, during the 2017 `WannaCry' ransomworm incident, it took 11 days from the disclosure of the vulnerability to the widespread distribution of the corresponding detection rules\cite{cisco01}.
Undisclosed zero-day vulnerabilities, which are primary targets for attackers, are typically not defensible with static rulesets\cite{enisa01}. 
Meanwhile, attack variants constructed using techniques such as adversarial machine learning\cite{alhajjar2021adversarial, zhang2022adversarial} further shorten rule lifespan and highlight the urgent need for timely updates to the ruleset\cite{google01, uetz2024you}.
This adversarial dynamic necessitates the automation of rapid rule generation and repair.

However, most existing studies\cite{du2021autocombo, alcantara2021syrius, zhang2020cmirgen, lee2016largen, coscia2024automatic, sohi2021rnnids} focus on the automatic generation of rules while neglecting the repair of existing ones. 
This oversight poses several critical challenges.
First, current methods struggle to determine the relationship between new attack samples and existing rules. 
Mainstream machine learning–based rule generation approaches typically rely on static feature extraction or pattern matching, thereby lacking effective integration of domain knowledge in cybersecurity. 
These models are consequently limited to surface-level matching and fail to capture the underlying semantics of attack behaviors, making it difficult to accurately assess the correspondence between samples and rules\cite{arp2022and, verkerken2022towards, apruzzese2022cross}.
Second, many existing methods tend to create new rules for every new sample, without mechanisms for repairing existing ones, which leads to severe rule redundancy\cite{liu2005complete,noiprasong2020ids}.
As a result, multiple rules correspond to different variants of the same attack. 
Such redundancy increases duplicate alerts, complicates rule management and deployment, and ultimately undermines system automation and maintainability\cite{vermeer2022ruling,vectra01,ho2012statistical}.
Recently, Large Language Models (LLMs), with their powerful contextual understanding and reasoning capabilities, have opened new possibilities for automatic rule generation and repair.
Leveraging their pretrained knowledge, LLMs can extract deep attack patterns from raw samples and thus hold the potential to produce high-quality rules. 
Nevertheless, the inherent hallucination problem of LLMs continues to hinder their practical adoption in this domain, as their outputs often fall short of the required precision and verifiability for rule deployment.
This limitation represents yet another major challenge for fully automated rule generation and repair.

To address these challenges, we propose a novel end-to-end framework, \textbf{GRIDAI}, for automatically generating and repairing intrusion detection rules through the collaboration of multiple LLM-based agents. 
Leveraging the reasoning capabilities of LLMs, this framework comprises four agents, each responsible for a specific task: determining the relationship between new samples and existing rules, generating new rules, repairing existing rules, and saving the rules to the rule repository.
Through agent collaboration, GRIDAI does not directly generate rules. 
Instead, it first uses an agent to assess the nature of incoming samples. 
If the sample represents a previously unseen attack type, the system forwards the relevant information to the rule-generation agent, which analyzes the attack details and generates a new rule.
If the sample is determined to be a variant of an attack already covered by an existing rule, the information is passed to another agent, which repairs the rule to enhance its generalization ability to handle variants.
The newly generated or updated rules are then stored by constructing or updating entries in the memory repository.
By incorporating LLMs, GRIDAI overcomes the limitations of traditional methods, thus enabling accurate inference of sample–rule relationships as well as effective rule generation and repair.
Additionally, GRIDAI integrates mechanisms such as real-time validation using the Model Context Protocol (MCP) tool, representative attack samples maintained for existing rules, and threshold-based error correction rollback, which effectively mitigate syntactic and semantic errors in rules caused by LLM hallucinations. 
Extensive experiments on two datasets demonstrate that GRIDAI can efficiently generate and repair intrusion detection rules to address newly discovered attacks and variants.

Our contributions are summarized as follows:
\begin{itemize}
    \item To the best of our knowledge, GRIDAI is the first system capable of both automatically generating and repairing intrusion detection rules, producing high-quality rules for novel and variant attacks, thereby enhancing detection capability against rapidly evolving threats.
    \item GRIDAI can accurately determine the relationship between new attack samples and existing rules, and automatically decide whether to generate or repair rules through LLM-based reasoning.
    \item GRIDAI effectively mitigates issues caused by LLM hallucinations and prevents erroneous model outputs from adversely affecting rule generation and repair through a set of error-correction mechanisms.
\end{itemize}

\section{Related Works}\label{sec:relat}
\subsection{Network Intrusion Detection Systems}
Among various Web attack detection methods, Network Intrusion Detection Systems (NIDS) play a pivotal role. 
They identify potential attack behaviors by analyzing network traffic and issue alerts to security administrators for timely response.
Based on detection paradigms, NIDS can be categorized into rule-based\cite{zhang2020cmirgen, alcantara2021syrius} and anomaly-based systems\cite{wang2021intrusion, du2023few}. 
Rule-based intrusion detection offers better interpretability and lower false-positive rates, which make it more widely adopted in real-world deployments\cite{sommer2010outside, tavallaee2010toward, vermeer2023alert}.
Moreover, it can directly process network traffic in real time without relying on specific hardware or software devices, which makes it easier to deploy and manage.
However, as cyberattack techniques continue to evolve, manually crafting detection rules has become increasingly impractical, highlighting the importance of automating rule generation to ensure timely and scalable defense.

\subsection{Automatic Rule Generation}\label{sec:rwarg}
As attack techniques continue to evolve, automating the construction of detection rules has become a central research challenge.
Existing data-mining–based signature extraction methods\cite{kim2004autograph, newsome2005polygraph, li2006hamsa, lee2016largen, zhang2020cmirgen, du2021autocombo, alcantara2021syrius} typically identify features common in malicious samples but rare in benign ones through tokenization or statistical analysis.
Another line of research combines anomaly detection with rule extraction by training classification models and deriving rules from their internal structures or decision boundaries\cite{hwang2007hybrid, yang2014combining, sohi2021rnnids, roldan2023automatic, coscia2024automatic, fan2025explainable}.
However, due to the reliance on surface-level statistical matching and lack of semantic understanding of attacks, most existing methods lack an effective mechanism to associate new samples with existing rules to guide repair.
Meanwhile, they focus only on rule generation while neglecting rule repair, leading to redundancy and inefficiency within rulesets\cite{vermeer2022ruling, vectra01, ho2012statistical}.
Therefore, GRIDAI employs LLMs incorporating cybersecurity knowledge to analyze incoming samples, determine their relationships with existing rules, and decide whether to generate new rules or repair existing ones, thereby enhancing the overall defense capability of RNIDS.

\subsection{LLMs for Cybersecurity}\label{sec:rwlfc}
Built upon the Transformer architecture and self-attention mechanisms\cite{vaswani2017attention}, Large Language Models (LLMs) learn patterns and contextual dependencies from massive datasets to generate fluent and coherent text.
In cybersecurity, researchers have constructed domain-specific knowledge datasets and evaluation benchmarks\cite{levi2025cyberpal} to adapt LLMs for security-oriented tasks.
LLMs have also been explored for machine-learning-based intrusion detection\cite{li2024ids}, API-misuse detection\cite{liu2024generating}, DDoS detection\cite{wang2024shieldgpt}, 1-day vulnerability detection\cite{fang2024llm}, shell-command interpretation\cite{deng2024raconteur}, and protocol fuzzing\cite{meng2024large}. 
For rule generation, LLMCloudHunter\cite{schwartz2025llmcloudhunter} leverages LLMs to automatically derive threat-detection rules from textual and visual open-source threat intelligence.
Other studies combine LLMs with security knowledge to generate NIDS rules\cite{du2025harnessing, hu2024llm}. 
We contend that, given their strong information-processing capabilities, LLMs can effectively compare new attack samples against existing rules and decide whether to generate or repair rules. 
Moreover, pervasive LLM hallucinations\cite{lin2021truthfulqa, huang2025survey} can introduce syntactic and semantic errors in the resulting rules, a limitation rarely considered in current cybersecurity tasks. 
One mechanism for mitigating hallucinations is interactive self-reflection\cite{ji2023towards}; however, the effectiveness of this self-correction strategy remains uncertain for more complex tasks.
Accordingly, GRIDAI introduces a multi-stage verification mechanism. Every rule operation must undergo real-time, tool-invoked validation. Upon detecting errors, the system rolls back and feeds the relevant information back to the LLMs, thereby improving the reliability of rule generation and repair.

\subsection{Multi-agent LLM Architecture}\label{sec:rwmla}
LLM agents can proactively invoke external tools to interact with their environment and dynamically accomplish tasks, giving them a distinct advantage over simple prompt-based approaches.
A multi-agent LLM architecture\cite{talebirad2023multi} is a collaborative system composed of multiple LLM-driven agents, each responsible for a specific subtask and coordinated through communication mechanisms.
Its core principle lies in task decomposition, role specialization, and information sharing, enabling multiple LLM agents to cooperate effectively to solve complex problems.
Such an architecture enhances system robustness, expands task-handling capacity, supports parallel execution, and provides greater flexibility and scalability.
Recent studies have demonstrated its remarkable performance across various complex tasks\cite{he2025llm, mao2025multi, wang2025cooperative, goswami2025chartcitor, ma2025local, qiao2025thematic, pei2025flow}.
Manual authoring of intrusion detection rules can be regarded as a multi-step process, in which tool invocation enables real-time rule validation and facilitates data management.
Therefore, GRIDAI is designed based on a multi-agent LLM architecture, leveraging its advantages to achieve dynamic rule generation and refinement.

\section{Methodology}\label{sec:methodology}

\begin{figure}[htb]
  \centering
  \includegraphics[width=\linewidth]{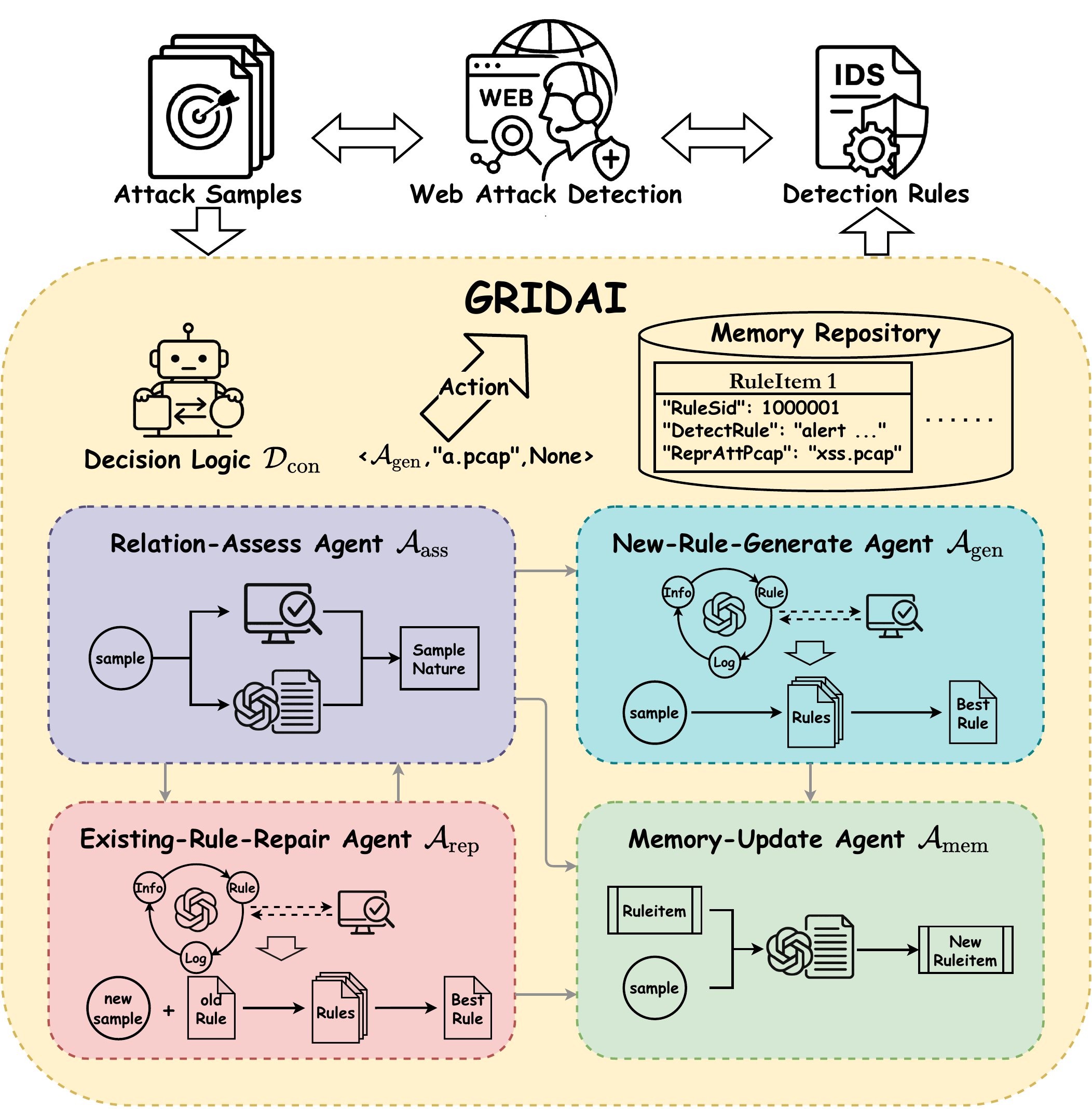}
    \caption{The Overview of GRIDAI.}
\label{fig:overview}
\end{figure}

\subsection{Problem Formulation}
Given a rule-based network intrusion detection system \( S \), GRIDAI aims to automatically generate and repair detection rules, constructing a deployable ruleset \( R \) from a series of attack traffic samples. 
Let \( P =\{p_{1},p_{2},\ldots,p_{n}\} \) denote the set of all attack samples, and \( R'=\{r_{1},r_{2},\ldots,r_{m}\} \) represent the set of existing rules in the rule memory repository.

GRIDAI processes the input samples one by one in a streaming manner.
For each sample \( p_i \in P \), GRIDAI first analyzes its nature and determines its relationship with existing rules in \( R' \):
\begin{itemize}
    \item \textbf{New attack type}: If \( p_i \) does not correspond to any attack type covered by existing rules, the system generates a new rule \( r_j \) and adds it to \( R' \) after validation by \(S\).
    \item \textbf{Attack variant}: If \( p_i \) is a variant of an attack type already covered by rule \( r_k \), the system repairs this existing rule and replaces it in \( R' \) with the newly generated rule \( r_k' \).
\end{itemize}

After all samples in \(P\) have been processed, GRIDAI outputs the final ruleset \( R' \) as the deployable ruleset \( R \), which provides excellent detection performance and exhibits strong generalization capabilities to handle new attack variants.

\subsection{Overview}\label{sec:overview}
As illustrated in Figure \ref{fig:overview}, GRIDAI is an end-to-end framework that leverages collaboration among multiple LLM-based agents for the automated generation and repair of intrusion detection rules.
According to subtask requirements, the agents in GRIDAI are categorized into four types: Relation-Assess agent \Aass, New-Rule-Generate agent \Agen, Existing-Rule-Repair agent \Arep, and Memory-Update agent \Amem.
Agents communicate and coordinate exclusively through action passing.
A fixed decision logic \Dcon\  schedules these agents as needed and routes actions between them, enabling the system to process incoming PCAP packets in a streaming fashion.
Additionally, we use a rule memory repository \Rmem\ to store and manage shared data among agents.

We employ a multi-agent system that decomposes the complex task into multiple subtasks, each executed by an independent agent module with its own short-term memory.
This design reduces the complexity and execution failure rate of individual subtasks, and prevents interference between different subtasks through role isolation.
Unlike simple prompt-based methods that interact with LLMs solely through APIs, the agent modules can invoke MCP tools to interact with the environment, for example, reading and writing to databases, checking rule formats, and testing rule effectiveness.
Building on this capability, we introduce two mechanisms to mitigate the impact of LLM hallucinations.
First, every new or repaired rule must undergo real-time simulated deployment and validation on NIDS. 
Second, if validation exceeds the preset failure threshold, the system rolls back the operation and feeds the relevant information back to the LLMs.
These mechanisms enable GRIDAI to intercept erroneous outputs, minimizing the influence of LLM hallucinations and ensuring that the generated rules are both syntactically and semantically correct.
Furthermore, the system maintains a representative attack sample for each rule. 
This ensures that even if an error occurs in classifying an attack sample under a certain rule, the system requires both the representative and new samples to pass validation, thereby causing the repair to fail and preventing error propagation.
This mechanism safeguards the final deployment results and preserves the rule generalization capability from being compromised.
Since each agent module's input and output are limited to a single action, GRIDAI can process different samples concurrently on multiple independent instances, supporting parallel execution and incremental input. 
Additionally, the system can preload a base set of rules into the rule memory repository to provide prior reference for future automated revisions.

\begin{figure*}[htb]
  \centering
  \includegraphics[width=\linewidth]{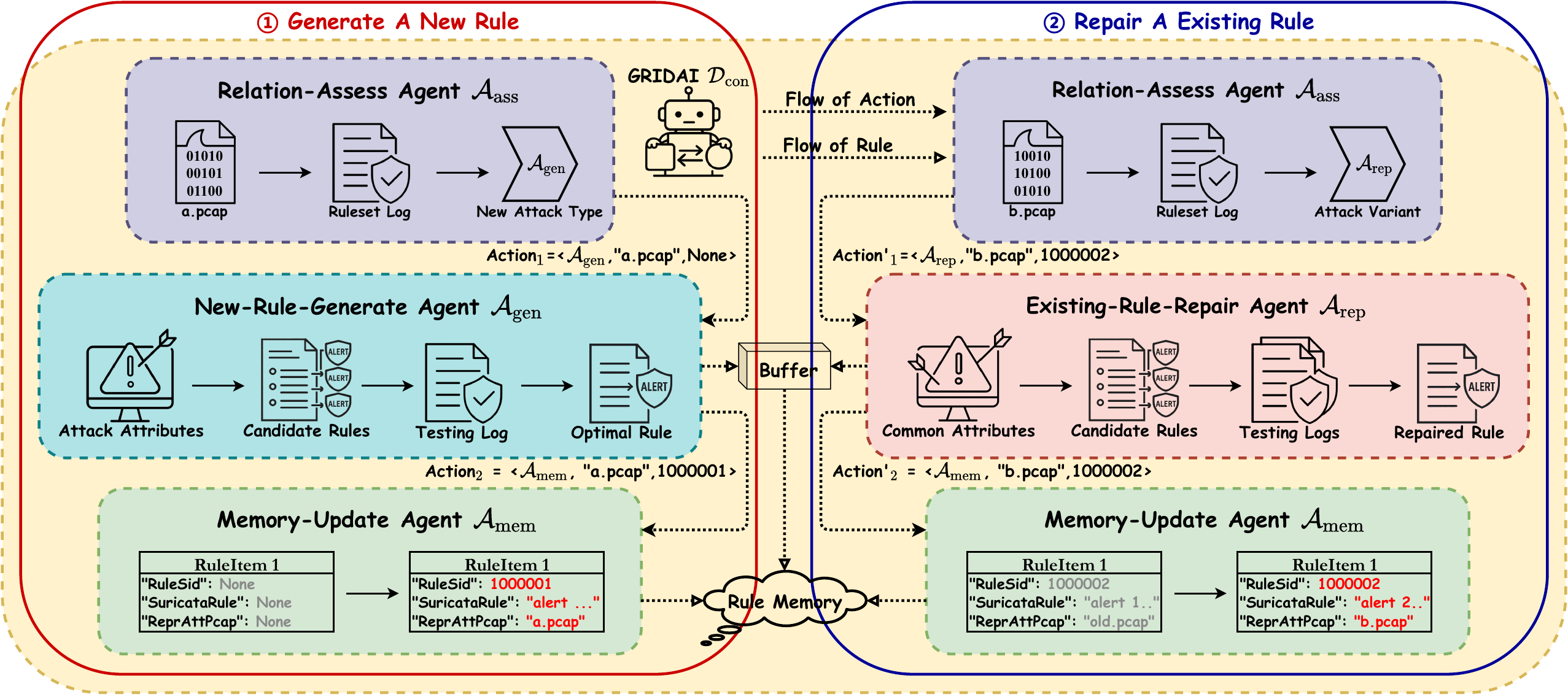}
    \caption{Examples of GRIDAI.}
\label{fig:examples}
\end{figure*}

\subsection{Components of GRIDAI}\label{sec:workflow}
\subsubsection{Decision Logic \Dcon\  for Action Management}
Decision Logic \Dcon\  orchestrates all atomic agents by managing the processing and transmission of actions.
Actions carry all the necessary information for operation, serving as the core of collaboration within the GRIDAI framework. 
Agent modules communicate indirectly via actions, meaning that both the input and output of each module are actions.
Formally, an action is defined as a triple:
$$A_i = <\mathcal{A},\  P,\  S>$$
where $\mathcal{A}\in \{\mathcal{A}_{ass},\ \mathcal{A}_{gen},\ \mathcal{A}_{rev},\ \mathcal{A}_{mem}\}$, $P$ is the identifier of the attack traffic, and $S \in \mathbb{N}\ \cup\ \{\bm{None}\}$ is the associated rule identifier.

\subsubsection{Rule Memory Repository \Rmem}
Rule Memory Repository \Rmem\ is another core component of GRIDAI, which is responsible for data holding as the long-time memory.
\Rmem\ is composed of multiple RuleItem entries, where each RuleItem corresponds to a single rule and is defined as a triple: $$RuleItem  = <RuleSid,\  DetectRule,\  ReprAttPcap>$$ where $RuleSid$ serves as the unique identifier used for indexing; $DetectRule$ field stores the rule whose sid equals $RuleSid$; and $ReprAttPcap$ contains a representative payload associated with the rule.
Through well-defined MCP tool interfaces, the LLM-based agent module can perform retrieval, insertion, and modification operations on RuleItems within \Rmem, while \Dcon \ can also manipulate the repository to retrieve the current ruleset as needed.
External rulesets may also be imported directly into the rule memory repository as a base to produce a more comprehensive ruleset and to improve protection against newly emerging attack variants.

\subsubsection{Relation-Assess Agent \Aass}
Acting as a network security expert in classifying attack behaviors, Relation-Assess Agent \Aass\ is the agent instantiated for each attack‐traffic sample upon its submission to GRIDAI. 
Its role is to choose the appropriate downstream agent for each PCAP. 
First, \Aass\  retrieves the current ruleset from \Rmem\  and launches the RNIDS engine for matching against the sample. 
If the engine produces an alert, it indicates that the existing ruleset is already effective. \Aass\ then records the identifier of the alerting rule in the action descriptor and forwards the sample to the Memory-Update Agent \Amem\ for subsequent processing.
If no alert is generated, \Aass\ extracts from \Rmem\  both the ruleset and metadata of representative attack samples, then combines them with the parsed payload of the current sample. 
This consolidated information is submitted to the LLM, which determines whether the new sample constitutes a variant of an existing attack behavior. 
Based on the LLM’s decision, \Aass\  dispatches the task either to \Agen\ or \Arep\  accordingly. 
In cases where the sample is determined to be a variant of an attack, the corresponding rule’s SID is passed as $S$ in the action.

\subsubsection{New-Rule-Generation Agent \Agen}
Acting as an IDS rule-authoring expert, New-Rule-Generation Agent \Agen\  designs a rule that produces a correct alert for the given novel sample.
Adopting a procedure like chain-of-thought\cite{wei2022chain}, we implement a two-stage rule generation pipeline.
First,\Agen\ instructs the LLM to thoroughly extract attack information from the payload (e.g., attack class, attack vector, and other relevant attributes).
Second, based on the extracted information, the LLM, which retains short-term memory from the previous step, generates multiple candidate IDS rules, yielding an initial candidate set.
A validation process follows: \Agen\  drives the detection engine to check each candidate’s format and test each rule against the attack sample.
If no candidate in a syntactically correct form produces a correct alert, \Agen\  returns the observed errors to the LLM and requests regeneration.
This loop repeats until a usable rule is produced. 
To avoid endless retries, we impose a retry threshold $M$: once exceeded, \Agen\  restarts the generation process.
When one or more candidates successfully trigger correct alerts, \Agen\  supplies their SIDs to the LLM and asks it to select the most effective rule.
The selected rule is forwarded to a rule buffer, and \Agen\  emits an action to \Amem\  to record the new rule into \Rmem.

\subsubsection{Existing-Rule-Repair Agent \Arep}
Acting as an IDS rule repair expert, Existing-Rule-Repair Agent \Arep\  repairs existing rules using the new attack traffic payload.
Following a similar prompt-and-validation process as \Agen, \Arep\  adapts existing rules to cover new attack variants.
Unlike new-rule generation, \Arep\  provides the LLM with the rule to be repaired, along with the rule’s representative payload and the new attack sample.
During detection-engine validation, a candidate repair is considered valid only if it produces correct alerts for both the new sample and the original one.
If the number of failed regeneration attempts for the candidate rule group exceeds $M$, we infer that \Aass\  may have misidentified the attack variant.
Instead of restarting \Arep\  as done in \Agen, the sample is returned to \Aass\  to restart the whole process.
If this cycle occurs more than a threshold $N$ times, we conclude that repairing existing rules cannot adequately cover the attack represented by the new sample.
The system therefore bypasses \Aass\  and forwards the sample directly to \Agen\  for new-rule generation.

\subsubsection{Memory‐Update Agent \Amem}
Memory‐Update Agent \Amem\ is responsible for updating the rule memory repository \Rmem\  when a rule is generated or repaired.
\Amem\  extracts the SID from the received action and queries \Rmem\  for the corresponding $RuleItem$; if no matching $RuleItem$ exists, \Amem\  creates a new one.
Next, \Amem\ retrieves the rule from the rule buffer and updates the $DetectRule$ field of the $RuleItem$.
Finally, if $ReprAttPcap$ already contains a representative attack sample, \Amem\  submits both the existing and new payloads, along with the rule, to the LLM to select the sample that best represents the attack behavior.
The selected sample is then saved into $ReprAttPcap$ and \Amem\ updates memory, marking the end of the current rule-generation or repair cycle.

\subsection{Workflow of GRIDAI}
As illustrated in Algorithm 1, the overall workflow of GRIDAI is orchestrated by Decision Logic \Dcon. 
Upon receiving a new attack sample, \Dcon\  initializes an action $A_0$ and dispatches it to \Aass\  for relation assessment. 
During execution, \Dcon\  dynamically instantiates agents as required and continuously monitors their operational states.
Whenever an agent outputs an action, \Dcon\  retrieves and routes it to the appropriate target agent according to the first element of the action tuple. 
Thresholds are also monitored in real time to detect repeated failures. 
If multiple rule-repair attempts fail to pass validation, the system infers that \Aass\  may have misidentified the attack variant and forwards the corresponding sample directly to \Agen\  for new-rule generation.
Once \Amem\  completes its update, the built-in validation mechanism guarantees that the rule memory repository \Rmem\  contains a rule capable of recognizing the attack represented by the processed sample, thereby marking the completion of the entire workflow for that sample.
  
\begin{algorithm}
\caption{Workflow of GRIDAI (Decision Logic \Dcon)}
\begin{algorithmic}[1]
    \State Initialize \Rmem = $\emptyset$;
    \For{$i$ = 1 to $n$}
        \State Initialize $count$ = 0 and obtain $p_{i}$ from $P$;
        \State Set $A_{i0}$ = <$\mathcal{A}_{ass},p_{i},\bm{None}$>
        \State Pass $A_{i0}$ into \Aass\ to determine the nature of $p_i$
        \State Obtain $A_{i1}$ = <$\mathcal{A}_1,p_i,s_1$> from \Aass;
        \If{$\mathcal{A}_1$ == \Agen\  \textbf{or} $count$ > $threshold$}
            \State Pass $A_{i1}$ into \Agen\ to generate a new rule $r_j$;
            \State Obtain $A_{i2}$ = <$\mathcal{A}_2,p_i,s_2$> from \Agen;
            \State Set $RuleItem_i  = <s_2,r_j,p_i>$;
            \State Pass $A_{i2}$ and $RuleItem_i$ into \Amem;
            \State Add a new rule in $R_{mem}$;
        \ElsIf{$\mathcal{A}_1$ == \Arep}
            \State Set $k$ = $s_1$;        
            \State Pass $A_{i1}$ into \Arep\ to repair the existing rule $r_k$;
            \State Obtain $A_{i3}$ = <$\mathcal{A}_3,p_i,s_1$> from \Arep;
            \If{$\mathcal{A}_3$ == \Aass}
                \State $count$ += 1;
                \State Jump to line 5 to analysis $p_i$ again;           
            \EndIf
            \State Set $RuleItem_i'  = <s_1,r_k',\bm{None}>$;
            \State Pass $A_{i3}$ and $RuleItem_i'$ into \Amem;
            \State Update the existing rule in $R_{mem}$;
        \EndIf
    \EndFor
\end{algorithmic}
\end{algorithm}

To provide a clearer illustration of this process, Figure \ref{fig:examples} presents two examples.
In Example 1, when a new attack sample ``a.pcap'' arrives, \Aass\  compares its attributes with existing rules stored in the repository. 
Because no sufficiently similar rule is found, the sample is identified as a new attack type and the action <\Agen,``a.pcap'',\textbf{None}> is created. \Agen\  then extracts the attack attributes, generates and tests multiple candidate rules, and selects the optimal detection rule. The resulting action <\Amem,``a.pcap'',1000001> triggers \Amem, which records this rule together with its representative sample ``a.pcap'' in \Rmem.
In Example 2, when another sample ``b.pcap'' is received, \Aass\ detects a strong similarity to the existing rule with sid ``1000002'' and identifies it as an attack variant. 
\Dcon\ passes the action <\Arep,``b.pcap'',1000002> to \Arep, which repairs the corresponding rule based on the common attributes between the new and previous samples. After validation, \Arep\  outputs <\Amem,``b.pcap'',1000002>, prompting \Amem\  to update \Rmem\  by replacing the old representative payload with ``b.pcap'' and storing the repaired rule.
Together, these examples demonstrate how GRIDAI generates and repairs intrusion detection rules, ensuring continuous adaptation to both novel attacks and evolving variants.

\section{Evaluation}\label{sec:exper}
GRIDAI automatically generates and repairs intrusion detection rules through the collaboration of multiple LLM-based agents. 
To comprehensively evaluate its performance, we design experiments focusing on the following key aspects: effectiveness, ablation, and sensitivity.

\textbf{Effectiveness:} Our first goal is to assess whether GRIDAI achieves significant improvements in rule generation and repair.
\begin{itemize}
    \item \emph{RQ1:} Can GRIDAI correctly determine the relationship between new attack samples and existing rules?
    To answer this question, we sequentially input novel attacks and their variants into the framework and observe whether GRIDAI correctly decides between generating new rules or repairing existing ones.
    \item \emph{RQ2:} How does the detection performance of rules produced by GRIDAI compare with that of traditional methods?
    We conduct a comprehensive evaluation across multiple metrics to verify whether GRIDAI achieves state-of-the-art performance.
\end{itemize}

\textbf{Ablation:} To further understand the contribution of each component within GRIDAI, we investigate the following research questions:
\begin{itemize}
    \item RQ3: How does the inclusion of the rule repair mechanism affect detection performance? 
    One of GRIDAI’s core innovations is its ability to modify existing rules in addition to generating new ones. 
    To validate its impact, we conduct an ablation study by removing the rule repair module \Arep\  and evaluate the resulting change in performance.
    \item RQ4: 
    How does the proposed hallucination mitigation mechanism influence GRIDAI's detection performance?
    To address hallucination-induced errors, GRIDAI performs syntactic and semantic validation after each rule operation and maintains a representative sample for every rule. 
    We conduct ablation experiments by removing these mechanisms separately to evaluate their individual contributions.
\end{itemize}

\textbf{Sensitivity:} Finally, we examine GRIDAI’s sensitivity to the choice of underlying LLM. In previous experiments, GPT-4.1 serves as the default model. Given the rapid evolution of LLMs, we further ask:
\begin{itemize}
    \item RQ5: How does GRIDAI’s performance change when GPT-4.1 is replaced with alternative LLMs, such as GLM-4-Plus or GPT-3.5?
    A well-designed LLM-based system should exhibit model-agnostic behavior—that is, remain effective even when the underlying model changes. This question thus examines GRIDAI’s robustness under different LLM configurations.
\end{itemize}

\subsection{Experimental Setup}
\subsubsection{Datesets.} 
In Web environments, network attacks are commonly launched through HTTP requests, making them one of the most prevalent and destructive forms of cyberattacks.
Typical examples include SQL injection, cross-site scripting (XSS), and cross-site request forgery (CSRF).
Therefore, we selected PCAP-format traffic packets containing HTTP-based Web attacks as the attack samples.
In addition, we used Suricata as the NIDS, which is a high-performance, open-source, and scalable network intrusion detection engine.

We first used the synthetic dataset provided in \cite{alcantara2021syrius}, selecting 14 samples corresponding to seven attack types as the training set and 54 samples as the test set. 
However, attack patterns in synthetic data often differ from those observed in real-world scenarios.
For instance, payloads in synthetic datasets are often considerably shorter. 
To address this gap, we additionally collected real-world data generated from a recent cyber defense exercise.
Using the latest ET ruleset\footnote{https://rules.emergingthreats.net/} (released on September 9, 2024) , we filtered attack samples from the captured traffic and obtained 75 training samples covering 43 attack types and 216 test samples.
Furthermore, when evaluating the quality of rulesets, false positives on benign traffic must also be considered, as excessive false alarms severely undermine its usability. 
Therefore, we collected over 18,000 benign samples from the same cyber defense exercise as part of the test set.
This design reflects the real-world distribution in Web environments, where normal traffic significantly outweighs attack traffic.

\subsubsection{Metrics.}
To comprehensively evaluate the performance of the ruleset, we used multiple metrics, including Total Alerts ($Alerts$), Detected Attacks ($DA$), Detection Rate ($DR$), Alert Duplication Rate ($ADR$), and Benign Alarm Rate ($BAR$). 
$Alerts$ refers to the total number of alerts triggered by the rules. 
An excessive number of alerts increases the analysis burden and can lead to alert fatigue. 
$DA$ represents the number of attack samples successfully identified by the rules. 
$DR$ measures the proportion of detected attack samples among all test attack samples, providing a direct reflection of the rule’s effectiveness. 
$ADR$ refers to the proportion of duplicate alerts triggered by attack samples. 
Similar to $Alerts$, a high $ADR$ indicates redundancy within the ruleset.
$BAR$ measures the proportion of benign traffic incorrectly triggering alerts. 
In practical rule deployment, false alarms are often the most critical metric.
In our ablation experiments, we further used Rule Count ($RC$), Rule Usability Rate ($RUR$), and Detection Latency ($DL$) to provide a more intuitive assessment of the ruleset's performance under different configurations.
$RC$ indicates the number of generated rules.
Ideally, the ruleset should remain as streamlined as possible while maintaining comparable detection performance.
$RUR$ represents the proportion of generated rules that pass format checks and are deployable. $DL$ measures the time required to complete detection across all test samples, which is crucial for real-time Web attack detection in practical scenarios.
An ideal ruleset should achieve efficient detection while maintaining a high detection rate, low false alarm rate, low redundancy, and streamlined structure.

\subsubsection{Baselines.}
To the best of our knowledge, no existing work has explicitly addressed automated rule repair.
Therefore, we selected three representative rule generation methods: AutoCombo\cite{du2021autocombo}, CMIRGen \cite{zhang2020cmirgen}, and Syrius \cite{alcantara2021syrius} as comparative baselines, all applicable to intrusion detection systems that directly take HTTP traffic as input.
Among them, AutoCombo and Syrius generate rules through data mining techniques, while CMIRGen derives detection signatures via clustering and model inference.
Notably, all three baselines require benign traffic as input, whereas GRIDAI only needs attack samples to generate rules.
Moreover, GRIDAI’s runtime depends solely on the response latency of the underlying LLMs, and its generated rules can be directly deployed without additional filtering. 
In contrast, AutoCombo and CMIRGen output detection signatures that require manual conversion into deployable rules, whereas Syrius directly outputs rules but relies on historical rulesets for heuristic ranking.
Consequently, a direct comparison of rule-generation speed between GRIDAI and these baselines is not practically meaningful.
Furthermore, we curated the corresponding rules from the ET ruleset for each attack sample in our experiments and used them as ground truth (GT) to comprehensively evaluate the detection performance of the generated rules.

\subsection{Effectiveness}
\subsubsection{RQ1: Sample–Rule Relationship Identification Accuracy.}
\begin{figure}[htb]
  \centering
  \includegraphics[width=\linewidth]{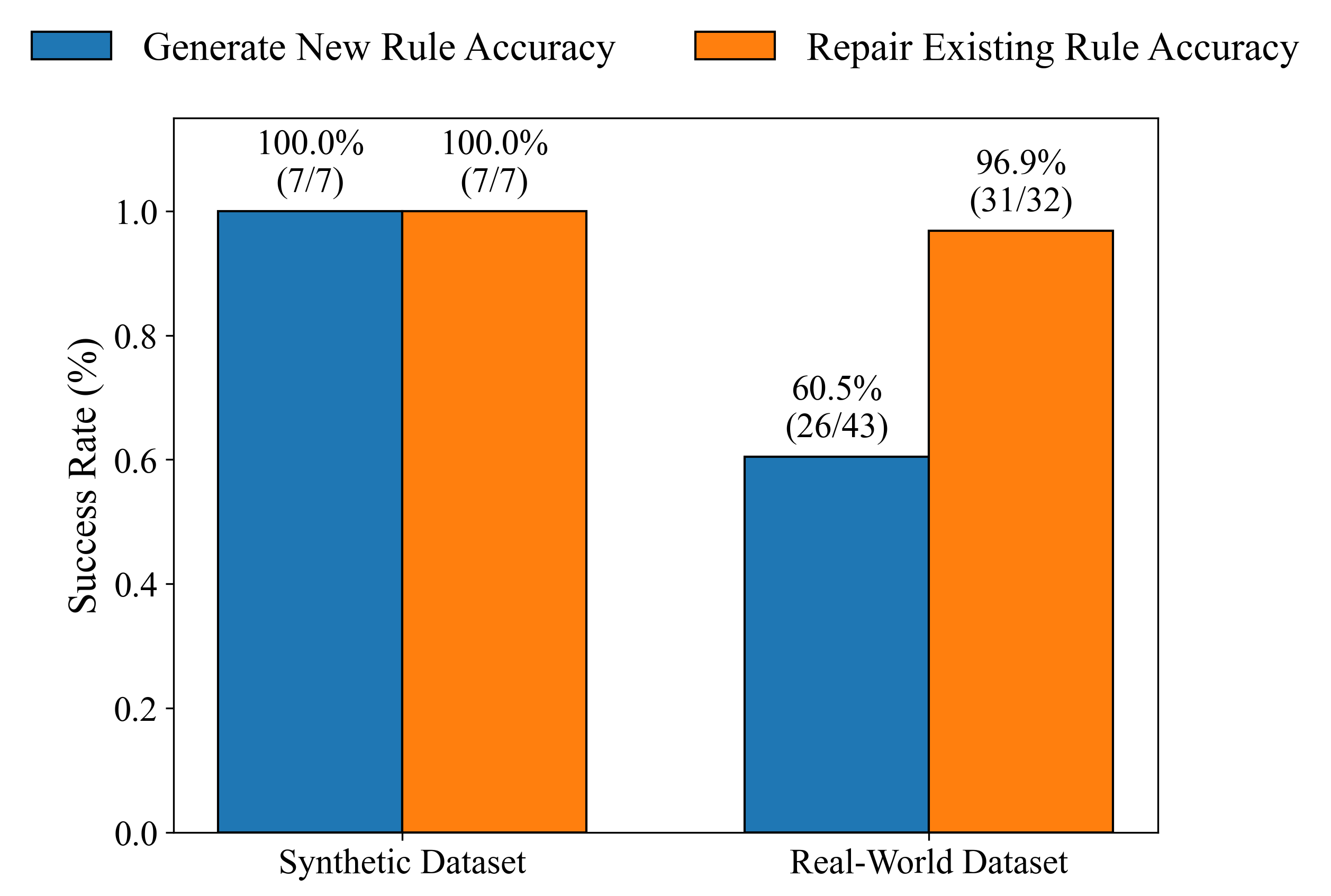}
    \caption{Accuracy of GRIDAI in Determining the Relationship between New Attack Samples and Existing Rules.}
\label{fig:RQ1}
\end{figure}

We recorded the input sequence of attack samples. 
For samples corresponding to the same ET rule, the first sample was labeled as one requiring new-rule generation, whereas subsequent samples were labeled as those expected to trigger rule repair.
The results were then compared with GRIDAI’s actual processing of the samples.
Figure \ref{fig:RQ1} illustrates the accuracy of GRIDAI's decisions regarding the relationships between new attack samples and existing rules.
On the synthetic dataset, GRIDAI achieved perfect accuracy in all decisions.
On the real-world dataset, GRIDAI made correct decisions in 96.9\% of cases requiring rule repair. 
Although the accuracy for new-rule generation was comparatively lower at 60.5\%, our dataset analysis indicated that this reduction was mainly due to multiple rules within the ET ruleset used for sample filtering sharing the same core detection fields.
For instance, the detection signatures of the rules with SIDs 2011768 and 2016415 are both ``<?php'', although they target different detection scenarios. 
GRIDAI generated a single rule capable of detecting both types of attacks, which to some extent demonstrates its robust judgment capability.

\subsubsection{RQ2: Detection Performance Comparison with Baselines.}
\begin{table*}[t]
   \small
   \caption{Performance Comparison of GRIDAI with Baselines}
   \label{tab:RQ2}
   \begin{tabular}{ccccccccccccc}
     \toprule
     \multirow{2}{*}{Methods} & \multicolumn{5}{c}{Synthetic Dataset} && \multicolumn{5}{c}{Real-World Dataset}\\
     \cline{2-6}\cline{8-12}
      & $Alerts$ $\downarrow$ & $DA$ $\uparrow$ & $DR$ $\uparrow$ & $ADR$ $\uparrow$ & $BAR$ $\downarrow$ && $Alerts$ $\downarrow$ & $DA$ $\uparrow$ & $DR$ $\uparrow$ & $ADR$ $\uparrow$ & $BAR$ $\downarrow$ \\
     \midrule
     AutoCombo\cite{du2021autocombo} & 16 & 16 & 0.296 & 0.000 & 0.000 && 23 & 23 & 0.106 & 0.000 & 0.000 \\
     CMIRGen\cite{zhang2020cmirgen} & 17 & 17 & 0.315 & 0.000 & 0.000 && 115 & 43 & 0.199 & 0.173 & 0.003\\
     Syrius\cite{alcantara2021syrius} & 1 & 1 & 0.019 & 0.000 & 0.000 & & 7 & 7 & 0.032 & 0.000 & 0.000 \\
     GT & 54 & 54 & 1.000 & 0.000 & 0.000 && 216 & 216 & 1.000 & 0.000 & 0.000\\
     GRIDAI & 54 & 54 & 1.000 & 0.000 & 0.000 && 236 & 206 & 0.954 & 0.127 & 0.000\\
     \bottomrule
   \end{tabular}
\end{table*}

\begin{table*}[t]
   \small
   \caption{Performance Comparison of GRIDAI with and without Key Modules}
   \label{tab:RQ3&4}
   \begin{tabular}{cccccccccccc}
     \toprule
     \multicolumn{3}{c}{Mechanism} && \multicolumn{8}{c}{Real-World Dataset}\\
     \cline{1-3}\cline{5-12}
     Repair & Tool & Sample && $Alerts$ $\downarrow$ & $DA$ $\uparrow$ & $DR$ $\uparrow$ & $ADR$ $\uparrow$ & $BAR$ $\downarrow$ & $RC$ $\downarrow$ & $RUR$ $\uparrow$ & $D$ $\uparrow$ \\
     \midrule
     \ding{51} & \ding{51} & \ding{51} && 236 & 206 & 0.954 & 0.127 & 0.000 & 27 & 1.000 & 30.122s \\
     \ding{55} & \ding{51} & \ding{51} && 885 & 204 & 0.944 & 0.668 & 0.014 & 73 & 1.000 & 33.986s\\
     \ding{51} & \ding{55} & \ding{51} && 180 & 156 & 0.722 & 0.128 & 0.001 & 43 & 0.860 & 30.063s\\
     \ding{51} & \ding{51} & \ding{55} && 383 & 187 & 0.866 & 0.142 & 0.009 & 47 & 1.000 & 36.232s\\
     \bottomrule
   \end{tabular}
\end{table*}
\begin{table*}[t]
   \small
   \caption{Performance Comparison of GRIDAI with Different LLMs}
   \label{tab:RQ5}
   \begin{tabular}{cccccccccccc}
     \toprule
     \multirow{2}{*}{LLMs} & \multicolumn{5}{c}{Synthetic Dataset} & & \multicolumn{5}{c}{Real-World Dataset}\\
     \cline{2-6}\cline{8-12}
      & $Alerts$ $\downarrow$ & $DA$ $\uparrow$ & $DR$ $\uparrow$ & $ADR$ $\uparrow$ & $BAR$ $\downarrow$ && $Alerts$ $\downarrow$ & $DA$ $\uparrow$ & $DR$ $\uparrow$ & $ADR$ $\uparrow$ & $BAR$ $\downarrow$ \\
     \midrule
     GPT-4.1 & 54 & 54 & 1.000 & 0.000 & 0.000 && 236 & 206 & 0.954 & 0.127 & 0.000\\
     GPT-4o & 54 & 54 & 1.000 & 0.000 & 0.000 && 310 & 198 & 0.917 & 0.236 & 0.003 \\
     GPT-3.5-turbo & 53 & 53 & 0.981 & 0.000 & 0.000 && 347 & 197 & 0.912 & 0.262 & 0.004 \\
     GLM-4-Plus & 54 & 54 & 1.000 & 0.000 & 0.000 && 280 & 203 & 0.940 & 0.262 & 0.001 \\
     GLM-4-Flash & 53 & 53 & 0.981 & 0.000 & 0.000 && 299 & 187 & 0.866 & 0.180 & 0.004\\
     \bottomrule
   \end{tabular}
\end{table*}

Table \ref{tab:RQ2} presents the performance metrics of detection rules generated by GRIDAI compared with the baselines. 
The results demonstrate that GRIDAI significantly outperforms the three baseline methods, achieving performance on the synthetic dataset comparable to that of manually crafted rulesets.
On the real-world dataset, the rules generated by GRIDAI successfully detected 95.4\% of attack samples, with an ADR of 12.7\% and zero false positives. 
Considering that duplicate-alert elimination techniques can be applied to reduce repeated alerts, while false positives remain the primary source of operational pressure in cybersecurity systems, this result is entirely acceptable. 
This indicates that GRIDAI possesses strong adaptability and practical value, even in more complex real-world scenarios.
In contrast, the three baseline methods performed poorly, primarily due to the extremely limited number of attack samples available for training. 
For each attack type, no more than four samples were used, reflecting the practical difficulty of collecting sufficient attack data in real-world settings.
The scarcity of samples caused the baselines to generate overly complex rules with limited generalization capability, resulting in frequent failures to trigger correct alerts.
On the real-world dataset, CMIRGen captured part of the attack behavior correctly but simultaneously generated a large number of redundant alerts and false positives.
It is also noteworthy that the baselines require benign samples during training and the manual grouping of attack samples according to behavioral variants. 
In contrast, by leveraging the pretrained knowledge and reasoning capabilities of LLMs, GRIDAI offers greater flexibility: it neither depends on benign samples nor requires attack traffic to be provided in a specific order or grouped by variant. 
The rules generated by GRIDAI are directly deployable without extensive manual inspection.

\subsection{Ablation}
\subsubsection{RQ3: Impact of the Rule Repair Mechanism on Detection Performance.}
We removed the Existing-Rule-Repair agent \Arep, causing GRIDAI to generate a new rule for each incoming attack sample.
Since the synthetic dataset consisted of relatively simple samples, its performance was barely affected by the ablation; however, real-world scenarios involve considerably more complex factors. 
To this end, Table \ref{tab:RQ3&4} reports the performance of the generated rules on the real-world dataset after removing \Arep. 
As shown, detection effectiveness showed no improvement, whereas the ADR increased sharply from 12.7\% to 66.8\%, accompanied by 1.4\% false positives and a 12.8\% increase in processing time.
These results demonstrate that the rule-repair mechanism is critical for GRIDAI to generate high-quality rulesets.
In practical security operations, both redundant alerts and false positives must be strictly controlled, as they can significantly reduce the efficiency of alert analysis; moreover, attack detection in Web environments is subject to stringent real-time requirements. 
Furthermore, the comparative results between GRIDAI and the baselines lacking a rule-repair mechanism in RQ2 indirectly confirm the importance of the rule-repair mechanism.

\subsubsection{RQ4: Impact of the Hallucination Mitigation Mechanism on Detection Performance.}
Real-time verification via tool invocation and representative attack samples are two critical mechanisms designed to mitigate the effects of LLM hallucinations. 
We conducted ablation studies on these mechanisms to evaluate their impact on GRIDAI. 
The third and fourth rows of Table \ref{tab:RQ3&4} present the experimental results on the real-world dataset. 
When the real-time verification mechanism was removed, the DR dropped from 95.4\% to 72.2\%, and the proportion of rules passing the NIDS format check decreased to 86.0\%. 
Under these conditions, some generated rules contained illegal fields in the Suricata rule format (e.g., the keyword ``is''), highlighting the crucial role of this mechanism in reducing syntactic and semantic errors caused by LLM hallucinations. 
Furthermore, removing the representative attack sample mechanism led to a substantial decline in detection performance, and the accuracy of GRIDAI in repairing existing rules decreased from 96.9\% to 68.8\%. 
Therefore, this mechanism play a key role in enabling GRIDAI to accurately determine the relationship between new samples and existing rules, while also working in conjunction with the verification mechanism to prevent excessive optimization of existing rules.

\subsection{Sensitivity}
\subsubsection{RQ5: Performance Sensitivity to Different underlying LLMs.}
We employed five LLMs from two different providers as the underlying models for GRIDAI. 
As shown in Table \ref{tab/tab_rq5}, GRIDAI based on GPT-4.1 produces the most effective detection rules, while other models achieve comparatively lower performance. 
This superiority can be attributed to GPT-4.1’s stronger reasoning and pattern-discovery capabilities, along with its enhanced knowledge integration.
Nevertheless, the rules generated by other models still hold practical value, particularly for lightweight or cost-sensitive deployments using open-access models such as GLM-4-Flash. 
In scenarios where data confidentiality is paramount, users of GRIDAI may also replace LLMs with privately hosted models to ensure higher security.
Importantly, thanks to its modular and model-agnostic design, the effectiveness of GRIDAI does not depend on any specific LLM.

\section{Conclusion}\label{sec:concl}
In this paper, we propose GRIDAI, an end-to-end framework that automatically generates and repairs intrusion detection rules.
It leverages the reasoning capabilities of LLMs through four specialized agents that collaborate to assess, generate, and repair rules. 
Unlike conventional methods, GRIDAI first classifies each incoming sample and dynamically dispatches agents for rule generation or repair. 
A tool-based real-time verification mechanism and representative attack samples further mitigate syntax and semantic errors caused by LLM hallucinations, enabling a fully automated workflow from samples to deployable rulesets.

Comprehensive experiments on two datasets demonstrate the effectiveness of GRIDAI. 
GRIDAI introduces a new paradigm for Web attack detection by incorporating LLM-driven rule generation and repair, enabling adaptive and continuously evolving rulesets without modifying existing systems.
Its hallucination-mitigation mechanisms offer generalizable insights for applying LLMs across cybersecurity tasks.
Further research will focus on extending this approach to diverse detection rule types and leveraging LLMs’ reasoning capabilities to conduct deeper analyses for generated rules, thereby offering richer insightes for Web security.

\bibliographystyle{ACM-Reference-Format}
\bibliography{refs}


\begin{thebibliography}{57}


\ifx \showCODEN    \undefined \def \showCODEN     #1{\unskip}     \fi
\ifx \showISBNx    \undefined \def \showISBNx     #1{\unskip}     \fi
\ifx \showISBNxiii \undefined \def \showISBNxiii  #1{\unskip}     \fi
\ifx \showISSN     \undefined \def \showISSN      #1{\unskip}     \fi
\ifx \showLCCN     \undefined \def \showLCCN      #1{\unskip}     \fi
\ifx \shownote     \undefined \def \shownote      #1{#1}          \fi
\ifx \showarticletitle \undefined \def \showarticletitle #1{#1}   \fi
\ifx \showURL      \undefined \def \showURL       {\relax}        \fi
\providecommand\bibfield[2]{#2}
\providecommand\bibinfo[2]{#2}
\providecommand\natexlab[1]{#1}
\providecommand\showeprint[2][]{arXiv:#2}

\bibitem[AI(2023)]%
        {vectra01}
\bibfield{author}{\bibinfo{person}{Vectra AI}.} \bibinfo{year}{2023}\natexlab{}.
\newblock \bibinfo{booktitle}{\emph{2023 State of Threat Detection}}.
\newblock
\urldef\tempurl%
\url{https://www.vectra.ai/resources/2023-state-of-threat-detection}
\showURL{%
Retrieved July 19, 2023 from \tempurl}


\bibitem[Alcantara et~al\mbox{.}(2021)]%
        {alcantara2021syrius}
\bibfield{author}{\bibinfo{person}{Lucas Alcantara}, \bibinfo{person}{Guilherme Padilha}, \bibinfo{person}{Rui Abreu}, {and} \bibinfo{person}{Marcelo d’Amorim}.} \bibinfo{year}{2021}\natexlab{}.
\newblock \showarticletitle{Syrius: Synthesis of rules for intrusion detectors}.
\newblock \bibinfo{journal}{\emph{IEEE Transactions on Reliability}} \bibinfo{volume}{71}, \bibinfo{number}{1} (\bibinfo{year}{2021}), \bibinfo{pages}{370--381}.
\newblock


\bibitem[Alhajjar et~al\mbox{.}(2021)]%
        {alhajjar2021adversarial}
\bibfield{author}{\bibinfo{person}{Elie Alhajjar}, \bibinfo{person}{Paul Maxwell}, {and} \bibinfo{person}{Nathaniel Bastian}.} \bibinfo{year}{2021}\natexlab{}.
\newblock \showarticletitle{Adversarial machine learning in network intrusion detection systems}.
\newblock \bibinfo{journal}{\emph{Expert Systems with Applications}}  \bibinfo{volume}{186} (\bibinfo{year}{2021}), \bibinfo{pages}{115782}.
\newblock


\bibitem[Apruzzese et~al\mbox{.}(2022)]%
        {apruzzese2022cross}
\bibfield{author}{\bibinfo{person}{Giovanni Apruzzese}, \bibinfo{person}{Luca Pajola}, {and} \bibinfo{person}{Mauro Conti}.} \bibinfo{year}{2022}\natexlab{}.
\newblock \showarticletitle{The cross-evaluation of machine learning-based network intrusion detection systems}.
\newblock \bibinfo{journal}{\emph{IEEE Transactions on Network and Service Management}} \bibinfo{volume}{19}, \bibinfo{number}{4} (\bibinfo{year}{2022}), \bibinfo{pages}{5152--5169}.
\newblock


\bibitem[Arp et~al\mbox{.}(2022)]%
        {arp2022and}
\bibfield{author}{\bibinfo{person}{Daniel Arp}, \bibinfo{person}{Erwin Quiring}, \bibinfo{person}{Feargus Pendlebury}, \bibinfo{person}{Alexander Warnecke}, \bibinfo{person}{Fabio Pierazzi}, \bibinfo{person}{Christian Wressnegger}, \bibinfo{person}{Lorenzo Cavallaro}, {and} \bibinfo{person}{Konrad Rieck}.} \bibinfo{year}{2022}\natexlab{}.
\newblock \showarticletitle{Dos and don'ts of machine learning in computer security}. In \bibinfo{booktitle}{\emph{31st USENIX Security Symposium (USENIX Security 22)}}. \bibinfo{pages}{3971--3988}.
\newblock


\bibitem[Coscia et~al\mbox{.}(2024)]%
        {coscia2024automatic}
\bibfield{author}{\bibinfo{person}{Antonio Coscia}, \bibinfo{person}{Vincenzo Dentamaro}, \bibinfo{person}{Stefano Galantucci}, \bibinfo{person}{Antonio Maci}, {and} \bibinfo{person}{Giuseppe Pirlo}.} \bibinfo{year}{2024}\natexlab{}.
\newblock \showarticletitle{Automatic decision tree-based NIDPS ruleset generation for DoS/DDoS attacks}.
\newblock \bibinfo{journal}{\emph{Journal of Information Security and Applications}}  \bibinfo{volume}{82} (\bibinfo{year}{2024}), \bibinfo{pages}{103736}.
\newblock


\bibitem[Deng et~al\mbox{.}(2024)]%
        {deng2024raconteur}
\bibfield{author}{\bibinfo{person}{Jiangyi Deng}, \bibinfo{person}{Xinfeng Li}, \bibinfo{person}{Yanjiao Chen}, \bibinfo{person}{Yijie Bai}, \bibinfo{person}{Haiqin Weng}, \bibinfo{person}{Yan Liu}, \bibinfo{person}{Tao Wei}, {and} \bibinfo{person}{Wenyuan Xu}.} \bibinfo{year}{2024}\natexlab{}.
\newblock \showarticletitle{Raconteur: A knowledgeable, insightful, and portable llm-powered shell command explainer}.
\newblock \bibinfo{journal}{\emph{arXiv preprint arXiv:2409.02074}} (\bibinfo{year}{2024}).
\newblock


\bibitem[D{\'\i}az-Verdejo et~al\mbox{.}(2022)]%
        {diaz2022detection}
\bibfield{author}{\bibinfo{person}{Jes{\'u}s D{\'\i}az-Verdejo}, \bibinfo{person}{Javier Mu{\~n}oz-Calle}, \bibinfo{person}{Antonio Estepa~Alonso}, \bibinfo{person}{Rafael Estepa~Alonso}, {and} \bibinfo{person}{Germ{\'a}n Madinabeitia}.} \bibinfo{year}{2022}\natexlab{}.
\newblock \showarticletitle{On the detection capabilities of signature-based intrusion detection systems in the context of web attacks}.
\newblock \bibinfo{journal}{\emph{Applied Sciences}} \bibinfo{volume}{12}, \bibinfo{number}{2} (\bibinfo{year}{2022}), \bibinfo{pages}{852}.
\newblock


\bibitem[Du et~al\mbox{.}(2023)]%
        {du2023few}
\bibfield{author}{\bibinfo{person}{Lei Du}, \bibinfo{person}{Zhaoquan Gu}, \bibinfo{person}{Ye Wang}, \bibinfo{person}{Le Wang}, {and} \bibinfo{person}{Yan Jia}.} \bibinfo{year}{2023}\natexlab{}.
\newblock \showarticletitle{A few-shot class-incremental learning method for network intrusion detection}.
\newblock \bibinfo{journal}{\emph{IEEE Transactions on Network and Service Management}} \bibinfo{volume}{21}, \bibinfo{number}{2} (\bibinfo{year}{2023}), \bibinfo{pages}{2389--2401}.
\newblock


\bibitem[Du et~al\mbox{.}(2025)]%
        {du2025harnessing}
\bibfield{author}{\bibinfo{person}{Lei Du}, \bibinfo{person}{Jiarui Li}, \bibinfo{person}{Hao Yan}, \bibinfo{person}{Yuhan Chai}, \bibinfo{person}{Binxing Fang}, {and} \bibinfo{person}{Zhaoquan Gu}.} \bibinfo{year}{2025}\natexlab{}.
\newblock \showarticletitle{Harnessing Large Language Models for Automated Intrusion Detection Rule Generation in Cyber Range}.
\newblock \bibinfo{journal}{\emph{IEEE Network}} (\bibinfo{year}{2025}).
\newblock


\bibitem[Du et~al\mbox{.}(2021)]%
        {du2021autocombo}
\bibfield{author}{\bibinfo{person}{Min Du}, \bibinfo{person}{Wenjun Hu}, {and} \bibinfo{person}{William Hewlett}.} \bibinfo{year}{2021}\natexlab{}.
\newblock \showarticletitle{Autocombo: Automatic malware signature generation through combination rule mining}. In \bibinfo{booktitle}{\emph{Proceedings of the 30th ACM International Conference on Information \& Knowledge Management}}. \bibinfo{pages}{3777--3786}.
\newblock


\bibitem[Fan et~al\mbox{.}(2025)]%
        {fan2025explainable}
\bibfield{author}{\bibinfo{person}{Mingrui Fan}, \bibinfo{person}{Jinxin Zuo}, \bibinfo{person}{Jiangwen Zhu}, {and} \bibinfo{person}{Yueming Lu}.} \bibinfo{year}{2025}\natexlab{}.
\newblock \showarticletitle{Explainable Anomaly-Based Intrusion Detection for Specialized IoT Environments Enabled by Rule Extraction from Autoencoder}.
\newblock \bibinfo{journal}{\emph{IEEE Internet of Things Journal}} (\bibinfo{year}{2025}).
\newblock


\bibitem[Fang et~al\mbox{.}(2024)]%
        {fang2024llm}
\bibfield{author}{\bibinfo{person}{Richard Fang}, \bibinfo{person}{Rohan Bindu}, \bibinfo{person}{Akul Gupta}, {and} \bibinfo{person}{Daniel Kang}.} \bibinfo{year}{2024}\natexlab{}.
\newblock \showarticletitle{Llm agents can autonomously exploit one-day vulnerabilities}.
\newblock \bibinfo{journal}{\emph{arXiv preprint arXiv:2404.08144}} (\bibinfo{year}{2024}).
\newblock


\bibitem[for Cybersecurity(2024)]%
        {enisa01}
\bibfield{author}{\bibinfo{person}{European Union~Agency for Cybersecurity}.} \bibinfo{year}{2024}\natexlab{}.
\newblock \bibinfo{booktitle}{\emph{ENISA Threat Landscape 2024}}.
\newblock
\urldef\tempurl%
\url{https://securitydelta.nl/media/com_hsd/report/690/document/ENISA-Threat-Landscape-2024.pdf}
\showURL{%
Retrieved September 19, 2024 from \tempurl}


\bibitem[Goswami et~al\mbox{.}(2025)]%
        {goswami2025chartcitor}
\bibfield{author}{\bibinfo{person}{Kanika Goswami}, \bibinfo{person}{Puneet Mathur}, \bibinfo{person}{Ryan Rossi}, {and} \bibinfo{person}{Franck Dernoncourt}.} \bibinfo{year}{2025}\natexlab{}.
\newblock \showarticletitle{ChartCitor: Answer Citations for ChartQA via Multi-Agent LLM Retrieval}. In \bibinfo{booktitle}{\emph{Companion Proceedings of the ACM on Web Conference 2025}}. \bibinfo{pages}{1668--1671}.
\newblock


\bibitem[He et~al\mbox{.}(2025)]%
        {he2025llm}
\bibfield{author}{\bibinfo{person}{Junda He}, \bibinfo{person}{Christoph Treude}, {and} \bibinfo{person}{David Lo}.} \bibinfo{year}{2025}\natexlab{}.
\newblock \showarticletitle{LLM-Based Multi-Agent Systems for Software Engineering: Literature Review, Vision, and the Road Ahead}.
\newblock \bibinfo{journal}{\emph{ACM Transactions on Software Engineering and Methodology}} \bibinfo{volume}{34}, \bibinfo{number}{5} (\bibinfo{year}{2025}), \bibinfo{pages}{1--30}.
\newblock


\bibitem[Ho et~al\mbox{.}(2012)]%
        {ho2012statistical}
\bibfield{author}{\bibinfo{person}{Cheng-Yuan Ho}, \bibinfo{person}{Yuan-Cheng Lai}, \bibinfo{person}{I-Wei Chen}, \bibinfo{person}{Fu-Yu Wang}, {and} \bibinfo{person}{Wei-Hsuan Tai}.} \bibinfo{year}{2012}\natexlab{}.
\newblock \showarticletitle{Statistical analysis of false positives and false negatives from real traffic with intrusion detection/prevention systems}.
\newblock \bibinfo{journal}{\emph{IEEE Communications Magazine}} \bibinfo{volume}{50}, \bibinfo{number}{3} (\bibinfo{year}{2012}), \bibinfo{pages}{146--154}.
\newblock


\bibitem[Hu et~al\mbox{.}(2024)]%
        {hu2024llm}
\bibfield{author}{\bibinfo{person}{Xiaowei Hu}, \bibinfo{person}{Haoning Chen}, \bibinfo{person}{Huaifeng Bao}, \bibinfo{person}{Wen Wang}, \bibinfo{person}{Feng Liu}, \bibinfo{person}{Guoqiao Zhou}, {and} \bibinfo{person}{Peng Yin}.} \bibinfo{year}{2024}\natexlab{}.
\newblock \showarticletitle{A LLM-based agent for the automatic generation and generalization of IDS rules}. In \bibinfo{booktitle}{\emph{2024 IEEE 23rd International Conference on Trust, Security and Privacy in Computing and Communications (TrustCom)}}. IEEE, \bibinfo{pages}{1875--1880}.
\newblock


\bibitem[Huang et~al\mbox{.}(2025)]%
        {huang2025survey}
\bibfield{author}{\bibinfo{person}{Lei Huang}, \bibinfo{person}{Weijiang Yu}, \bibinfo{person}{Weitao Ma}, \bibinfo{person}{Weihong Zhong}, \bibinfo{person}{Zhangyin Feng}, \bibinfo{person}{Haotian Wang}, \bibinfo{person}{Qianglong Chen}, \bibinfo{person}{Weihua Peng}, \bibinfo{person}{Xiaocheng Feng}, \bibinfo{person}{Bing Qin}, {et~al\mbox{.}}} \bibinfo{year}{2025}\natexlab{}.
\newblock \showarticletitle{A survey on hallucination in large language models: Principles, taxonomy, challenges, and open questions}.
\newblock \bibinfo{journal}{\emph{ACM Transactions on Information Systems}} \bibinfo{volume}{43}, \bibinfo{number}{2} (\bibinfo{year}{2025}), \bibinfo{pages}{1--55}.
\newblock


\bibitem[Hwang et~al\mbox{.}(2007)]%
        {hwang2007hybrid}
\bibfield{author}{\bibinfo{person}{Kai Hwang}, \bibinfo{person}{Min Cai}, \bibinfo{person}{Ying Chen}, {and} \bibinfo{person}{Min Qin}.} \bibinfo{year}{2007}\natexlab{}.
\newblock \showarticletitle{Hybrid intrusion detection with weighted signature generation over anomalous internet episodes}.
\newblock \bibinfo{journal}{\emph{IEEE Transactions on dependable and secure computing}} \bibinfo{volume}{4}, \bibinfo{number}{1} (\bibinfo{year}{2007}), \bibinfo{pages}{41--55}.
\newblock


\bibitem[Ji et~al\mbox{.}(2023)]%
        {ji2023towards}
\bibfield{author}{\bibinfo{person}{Ziwei Ji}, \bibinfo{person}{Tiezheng Yu}, \bibinfo{person}{Yan Xu}, \bibinfo{person}{Nayeon Lee}, \bibinfo{person}{Etsuko Ishii}, {and} \bibinfo{person}{Pascale Fung}.} \bibinfo{year}{2023}\natexlab{}.
\newblock \showarticletitle{Towards mitigating LLM hallucination via self reflection}. In \bibinfo{booktitle}{\emph{Findings of the Association for Computational Linguistics: EMNLP 2023}}. \bibinfo{pages}{1827--1843}.
\newblock


\bibitem[Kim and Karp(2004)]%
        {kim2004autograph}
\bibfield{author}{\bibinfo{person}{Hyang-Ah Kim} {and} \bibinfo{person}{Brad Karp}.} \bibinfo{year}{2004}\natexlab{}.
\newblock \showarticletitle{Autograph: Toward Automated, Distributed Worm Signature Detection.}. In \bibinfo{booktitle}{\emph{USENIX security symposium}}, Vol.~\bibinfo{volume}{286}. San Diego, CA.
\newblock


\bibitem[Lee et~al\mbox{.}(2016)]%
        {lee2016largen}
\bibfield{author}{\bibinfo{person}{Suchul Lee}, \bibinfo{person}{Sungho Kim}, \bibinfo{person}{Sungil Lee}, \bibinfo{person}{Jaehyuk Choi}, \bibinfo{person}{Hanjun Yoon}, \bibinfo{person}{Dohoon Lee}, {and} \bibinfo{person}{Jun-Rak Lee}.} \bibinfo{year}{2016}\natexlab{}.
\newblock \showarticletitle{LARGen: automatic signature generation for Malwares using latent Dirichlet allocation}.
\newblock \bibinfo{journal}{\emph{IEEE Transactions on Dependable and Secure Computing}} \bibinfo{volume}{15}, \bibinfo{number}{5} (\bibinfo{year}{2016}), \bibinfo{pages}{771--783}.
\newblock


\bibitem[Levi et~al\mbox{.}(2025)]%
        {levi2025cyberpal}
\bibfield{author}{\bibinfo{person}{Matan Levi}, \bibinfo{person}{Yair Allouche}, \bibinfo{person}{Daniel Ohayon}, {and} \bibinfo{person}{Anton Puzanov}.} \bibinfo{year}{2025}\natexlab{}.
\newblock \showarticletitle{Cyberpal. ai: Empowering llms with expert-driven cybersecurity instructions}. In \bibinfo{booktitle}{\emph{Proceedings of the AAAI Conference on Artificial Intelligence}}, Vol.~\bibinfo{volume}{39}. \bibinfo{pages}{24402--24412}.
\newblock


\bibitem[Li et~al\mbox{.}(2024)]%
        {li2024ids}
\bibfield{author}{\bibinfo{person}{Yanjie Li}, \bibinfo{person}{Zhen Xiang}, \bibinfo{person}{Nathaniel~D Bastian}, \bibinfo{person}{Dawn Song}, {and} \bibinfo{person}{Bo Li}.} \bibinfo{year}{2024}\natexlab{}.
\newblock \showarticletitle{IDS-Agent: An LLM Agent for Explainable Intrusion Detection in IoT Networks}. In \bibinfo{booktitle}{\emph{NeurIPS 2024 Workshop on Open-World Agents}}.
\newblock


\bibitem[Li et~al\mbox{.}(2006)]%
        {li2006hamsa}
\bibfield{author}{\bibinfo{person}{Zhichun Li}, \bibinfo{person}{Manan Sanghi}, \bibinfo{person}{Yan Chen}, \bibinfo{person}{Ming-Yang Kao}, {and} \bibinfo{person}{Brian Chavez}.} \bibinfo{year}{2006}\natexlab{}.
\newblock \showarticletitle{Hamsa: Fast signature generation for zero-day polymorphic worms with provable attack resilience}. In \bibinfo{booktitle}{\emph{2006 IEEE Symposium on Security and Privacy (S\&P'06)}}. IEEE, \bibinfo{pages}{15--pp}.
\newblock


\bibitem[Lin et~al\mbox{.}(2021)]%
        {lin2021truthfulqa}
\bibfield{author}{\bibinfo{person}{Stephanie Lin}, \bibinfo{person}{Jacob Hilton}, {and} \bibinfo{person}{Owain Evans}.} \bibinfo{year}{2021}\natexlab{}.
\newblock \showarticletitle{Truthfulqa: Measuring how models mimic human falsehoods}.
\newblock \bibinfo{journal}{\emph{arXiv preprint arXiv:2109.07958}} (\bibinfo{year}{2021}).
\newblock


\bibitem[Liu and Gouda(2005)]%
        {liu2005complete}
\bibfield{author}{\bibinfo{person}{Alex~X Liu} {and} \bibinfo{person}{Mohamed~G Gouda}.} \bibinfo{year}{2005}\natexlab{}.
\newblock \showarticletitle{Complete redundancy detection in firewalls}. In \bibinfo{booktitle}{\emph{IFIP Annual Conference on Data and Applications Security and Privacy}}. Springer, \bibinfo{pages}{193--206}.
\newblock


\bibitem[Liu et~al\mbox{.}(2024)]%
        {liu2024generating}
\bibfield{author}{\bibinfo{person}{Jinghua Liu}, \bibinfo{person}{Yi Yang}, \bibinfo{person}{Kai Chen}, {and} \bibinfo{person}{Miaoqian Lin}.} \bibinfo{year}{2024}\natexlab{}.
\newblock \showarticletitle{Generating api parameter security rules with llm for api misuse detection}.
\newblock \bibinfo{journal}{\emph{arXiv preprint arXiv:2409.09288}} (\bibinfo{year}{2024}).
\newblock


\bibitem[Ma et~al\mbox{.}(2025)]%
        {ma2025local}
\bibfield{author}{\bibinfo{person}{Jiatong Ma}, \bibinfo{person}{Linmei Hu}, \bibinfo{person}{Rang Li}, {and} \bibinfo{person}{Wenbo Fu}.} \bibinfo{year}{2025}\natexlab{}.
\newblock \showarticletitle{Local: Logical and causal fact-checking with llm-based multi-agents}. In \bibinfo{booktitle}{\emph{Proceedings of the ACM on Web Conference 2025}}. \bibinfo{pages}{1614--1625}.
\newblock


\bibitem[Mao et~al\mbox{.}(2025)]%
        {mao2025multi}
\bibfield{author}{\bibinfo{person}{Teng Mao}, \bibinfo{person}{Shuangtao Yang}, {and} \bibinfo{person}{Bo Fu}.} \bibinfo{year}{2025}\natexlab{}.
\newblock \showarticletitle{A Multi-Agent Framework for Multi-Source Manufacturing Knowledge Integration and Question Answering}. In \bibinfo{booktitle}{\emph{Companion Proceedings of the ACM on Web Conference 2025}}. \bibinfo{pages}{1687--1695}.
\newblock


\bibitem[Meng et~al\mbox{.}(2024)]%
        {meng2024large}
\bibfield{author}{\bibinfo{person}{Ruijie Meng}, \bibinfo{person}{Martin Mirchev}, \bibinfo{person}{Marcel B{\"o}hme}, {and} \bibinfo{person}{Abhik Roychoudhury}.} \bibinfo{year}{2024}\natexlab{}.
\newblock \showarticletitle{Large language model guided protocol fuzzing}. In \bibinfo{booktitle}{\emph{Proceedings of the 31st Annual Network and Distributed System Security Symposium (NDSS)}}, Vol.~\bibinfo{volume}{2024}.
\newblock


\bibitem[Newsome et~al\mbox{.}(2005)]%
        {newsome2005polygraph}
\bibfield{author}{\bibinfo{person}{James Newsome}, \bibinfo{person}{Brad Karp}, {and} \bibinfo{person}{Dawn Song}.} \bibinfo{year}{2005}\natexlab{}.
\newblock \showarticletitle{Polygraph: Automatically generating signatures for polymorphic worms}. In \bibinfo{booktitle}{\emph{2005 IEEE Symposium on Security and Privacy (S\&P'05)}}. \bibinfo{publisher}{IEEE}, \bibinfo{pages}{226--241}.
\newblock


\bibitem[Noiprasong and Khurat(2020)]%
        {noiprasong2020ids}
\bibfield{author}{\bibinfo{person}{Piyawat Noiprasong} {and} \bibinfo{person}{Assadarat Khurat}.} \bibinfo{year}{2020}\natexlab{}.
\newblock \showarticletitle{An ids rule redundancy verification}. In \bibinfo{booktitle}{\emph{2020 17th International Joint Conference on Computer Science and Software Engineering (JCSSE)}}. IEEE, \bibinfo{pages}{110--115}.
\newblock


\bibitem[of~Investigation(2025)]%
        {fbi01}
\bibfield{author}{\bibinfo{person}{Federal~Bureau of Investigation}.} \bibinfo{year}{2025}\natexlab{}.
\newblock \bibinfo{booktitle}{\emph{Internet Crime Report 2024}}.
\newblock
\urldef\tempurl%
\url{https://www.ic3.gov/AnnualReport/Reports/2024_IC3Report.pdf}
\showURL{%
Retrieved April 23, 2025 from \tempurl}


\bibitem[Pei et~al\mbox{.}(2025)]%
        {pei2025flow}
\bibfield{author}{\bibinfo{person}{Changhua Pei}, \bibinfo{person}{Zexin Wang}, \bibinfo{person}{Fengrui Liu}, \bibinfo{person}{Zeyan Li}, \bibinfo{person}{Yang Liu}, \bibinfo{person}{Xiao He}, \bibinfo{person}{Rong Kang}, \bibinfo{person}{Tieying Zhang}, \bibinfo{person}{Jianjun Chen}, \bibinfo{person}{Jianhui Li}, {et~al\mbox{.}}} \bibinfo{year}{2025}\natexlab{}.
\newblock \showarticletitle{Flow-of-Action: SOP Enhanced LLM-Based Multi-Agent System for Root Cause Analysis}. In \bibinfo{booktitle}{\emph{Companion Proceedings of the ACM on Web Conference 2025}}. \bibinfo{pages}{422--431}.
\newblock


\bibitem[Qiao et~al\mbox{.}(2025)]%
        {qiao2025thematic}
\bibfield{author}{\bibinfo{person}{Tingrui Qiao}, \bibinfo{person}{Caroline Walker}, \bibinfo{person}{Chris Cunningham}, {and} \bibinfo{person}{Yun~Sing Koh}.} \bibinfo{year}{2025}\natexlab{}.
\newblock \showarticletitle{Thematic-LM: a LLM-based multi-agent system for large-scale thematic analysis}. In \bibinfo{booktitle}{\emph{Proceedings of the ACM on Web Conference 2025}}. \bibinfo{pages}{649--658}.
\newblock


\bibitem[Rold{\'a}n-G{\'o}mez et~al\mbox{.}(2023)]%
        {roldan2023automatic}
\bibfield{author}{\bibinfo{person}{Jos{\'e} Rold{\'a}n-G{\'o}mez}, \bibinfo{person}{Juan Boubeta-Puig}, \bibinfo{person}{Javier Carrillo-Mondejar}, \bibinfo{person}{Juan Manuel~Castelo Gomez}, {and} \bibinfo{person}{Jes{\'u}s~Mart{\'\i}nez del Rinc{\'o}n}.} \bibinfo{year}{2023}\natexlab{}.
\newblock \showarticletitle{An automatic complex event processing rules generation system for the recognition of real-time IoT attack patterns}.
\newblock \bibinfo{journal}{\emph{Engineering Applications of Artificial Intelligence}}  \bibinfo{volume}{123} (\bibinfo{year}{2023}), \bibinfo{pages}{106344}.
\newblock


\bibitem[Schwartz et~al\mbox{.}(2025)]%
        {schwartz2025llmcloudhunter}
\bibfield{author}{\bibinfo{person}{Yuval Schwartz}, \bibinfo{person}{Lavi Ben-Shimol}, \bibinfo{person}{Dudu Mimran}, \bibinfo{person}{Yuval Elovici}, {and} \bibinfo{person}{Asaf Shabtai}.} \bibinfo{year}{2025}\natexlab{}.
\newblock \showarticletitle{Llmcloudhunter: Harnessing llms for automated extraction of detection rules from cloud-based cti}. In \bibinfo{booktitle}{\emph{Proceedings of the ACM on Web Conference 2025}}. \bibinfo{pages}{1922--1941}.
\newblock


\bibitem[Security(2024)]%
        {google01}
\bibfield{author}{\bibinfo{person}{Google~Cloud Security}.} \bibinfo{year}{2024}\natexlab{}.
\newblock \bibinfo{booktitle}{\emph{M-trends 2024: Attacker Evasion Trends Report}}.
\newblock
\urldef\tempurl%
\url{https://services.google.com/fh/files/misc/m-trends-2024.pdf}
\showURL{%
Retrieved April 23, 2024 from \tempurl}


\bibitem[Sohi et~al\mbox{.}(2021)]%
        {sohi2021rnnids}
\bibfield{author}{\bibinfo{person}{Soroush~M Sohi}, \bibinfo{person}{Jean-Pierre Seifert}, {and} \bibinfo{person}{Fatemeh Ganji}.} \bibinfo{year}{2021}\natexlab{}.
\newblock \showarticletitle{RNNIDS: Enhancing network intrusion detection systems through deep learning}.
\newblock \bibinfo{journal}{\emph{Computers \& Security}}  \bibinfo{volume}{102} (\bibinfo{year}{2021}), \bibinfo{pages}{102151}.
\newblock


\bibitem[Sommer and Paxson(2010)]%
        {sommer2010outside}
\bibfield{author}{\bibinfo{person}{Robin Sommer} {and} \bibinfo{person}{Vern Paxson}.} \bibinfo{year}{2010}\natexlab{}.
\newblock \showarticletitle{Outside the closed world: On using machine learning for network intrusion detection}. In \bibinfo{booktitle}{\emph{2010 IEEE symposium on security and privacy}}. IEEE, \bibinfo{pages}{305--316}.
\newblock


\bibitem[Systems(2017)]%
        {cisco01}
\bibfield{author}{\bibinfo{person}{Cisco Systems}.} \bibinfo{year}{2017}\natexlab{}.
\newblock \bibinfo{booktitle}{\emph{WannaCry Ransomware Defense Timeline}}.
\newblock
\urldef\tempurl%
\url{https://www.cisco.com/c/dam/en/us/solutions/collateral/enterprise-networks/ransomware-defense/wannacry-timeline.pdf}
\showURL{%
Retrieved May 12, 2017 from \tempurl}


\bibitem[Talebirad and Nadiri(2023)]%
        {talebirad2023multi}
\bibfield{author}{\bibinfo{person}{Yashar Talebirad} {and} \bibinfo{person}{Amirhossein Nadiri}.} \bibinfo{year}{2023}\natexlab{}.
\newblock \showarticletitle{Multi-agent collaboration: Harnessing the power of intelligent llm agents}.
\newblock \bibinfo{journal}{\emph{arXiv preprint arXiv:2306.03314}} (\bibinfo{year}{2023}).
\newblock


\bibitem[Tavallaee et~al\mbox{.}(2010)]%
        {tavallaee2010toward}
\bibfield{author}{\bibinfo{person}{Mahbod Tavallaee}, \bibinfo{person}{Natalia Stakhanova}, {and} \bibinfo{person}{Ali~Akbar Ghorbani}.} \bibinfo{year}{2010}\natexlab{}.
\newblock \showarticletitle{Toward credible evaluation of anomaly-based intrusion-detection methods}.
\newblock \bibinfo{journal}{\emph{IEEE Transactions on Systems, Man, and Cybernetics, Part C (Applications and Reviews)}} \bibinfo{volume}{40}, \bibinfo{number}{5} (\bibinfo{year}{2010}), \bibinfo{pages}{516--524}.
\newblock


\bibitem[Uetz et~al\mbox{.}(2024)]%
        {uetz2024you}
\bibfield{author}{\bibinfo{person}{Rafael Uetz}, \bibinfo{person}{Marco Herzog}, \bibinfo{person}{Louis Hackl{\"a}nder}, \bibinfo{person}{Simon Schwarz}, {and} \bibinfo{person}{Martin Henze}.} \bibinfo{year}{2024}\natexlab{}.
\newblock \showarticletitle{You Cannot Escape Me: Detecting Evasions of $\{$SIEM$\}$ Rules in Enterprise Networks}. In \bibinfo{booktitle}{\emph{33rd USENIX Security Symposium (USENIX Security 24)}}. \bibinfo{publisher}{USENIX Association}, \bibinfo{pages}{5179--5196}.
\newblock


\bibitem[Vaswani et~al\mbox{.}(2017)]%
        {vaswani2017attention}
\bibfield{author}{\bibinfo{person}{Ashish Vaswani}, \bibinfo{person}{Noam Shazeer}, \bibinfo{person}{Niki Parmar}, \bibinfo{person}{Jakob Uszkoreit}, \bibinfo{person}{Llion Jones}, \bibinfo{person}{Aidan~N Gomez}, \bibinfo{person}{Lukasz Kaiser}, {and} \bibinfo{person}{Illia Polosukhin}.} \bibinfo{year}{2017}\natexlab{}.
\newblock \showarticletitle{Attention is all you need}.
\newblock \bibinfo{journal}{\emph{Advances in neural information processing systems}}  \bibinfo{volume}{30} (\bibinfo{year}{2017}).
\newblock


\bibitem[Verkerken et~al\mbox{.}(2022)]%
        {verkerken2022towards}
\bibfield{author}{\bibinfo{person}{Miel Verkerken}, \bibinfo{person}{Laurens D’hooge}, \bibinfo{person}{Tim Wauters}, \bibinfo{person}{Bruno Volckaert}, {and} \bibinfo{person}{Filip De~Turck}.} \bibinfo{year}{2022}\natexlab{}.
\newblock \showarticletitle{Towards model generalization for intrusion detection: Unsupervised machine learning techniques}.
\newblock \bibinfo{journal}{\emph{Journal of Network and Systems Management}} \bibinfo{volume}{30}, \bibinfo{number}{1} (\bibinfo{year}{2022}), \bibinfo{pages}{12}.
\newblock


\bibitem[Vermeer et~al\mbox{.}(2023)]%
        {vermeer2023alert}
\bibfield{author}{\bibinfo{person}{Mathew Vermeer}, \bibinfo{person}{Natalia Kadenko}, \bibinfo{person}{Michel van Eeten}, \bibinfo{person}{Carlos Ga{\~n}{\'a}n}, {and} \bibinfo{person}{Simon Parkin}.} \bibinfo{year}{2023}\natexlab{}.
\newblock \showarticletitle{Alert alchemy: SOC workflows and decisions in the management of NIDS rules}. In \bibinfo{booktitle}{\emph{Proceedings of the 2023 ACM SIGSAC Conference on Computer and Communications Security}}. \bibinfo{pages}{2770--2784}.
\newblock


\bibitem[Vermeer et~al\mbox{.}(2022)]%
        {vermeer2022ruling}
\bibfield{author}{\bibinfo{person}{Mathew Vermeer}, \bibinfo{person}{Michel Van~Eeten}, {and} \bibinfo{person}{Carlos Ga{\~n}{\'a}n}.} \bibinfo{year}{2022}\natexlab{}.
\newblock \showarticletitle{Ruling the rules: Quantifying the evolution of rulesets, alerts and incidents in network intrusion detection}. In \bibinfo{booktitle}{\emph{Proceedings of the 2022 ACM on Asia Conference on Computer and Communications Security}}. \bibinfo{pages}{799--814}.
\newblock


\bibitem[Wang et~al\mbox{.}(2024)]%
        {wang2024shieldgpt}
\bibfield{author}{\bibinfo{person}{Tongze Wang}, \bibinfo{person}{Xiaohui Xie}, \bibinfo{person}{Lei Zhang}, \bibinfo{person}{Chuyi Wang}, \bibinfo{person}{Liang Zhang}, {and} \bibinfo{person}{Yong Cui}.} \bibinfo{year}{2024}\natexlab{}.
\newblock \showarticletitle{Shieldgpt: An llm-based framework for ddos mitigation}. In \bibinfo{booktitle}{\emph{Proceedings of the 8th asia-pacific workshop on networking}}. \bibinfo{pages}{108--114}.
\newblock


\bibitem[Wang et~al\mbox{.}(2021)]%
        {wang2021intrusion}
\bibfield{author}{\bibinfo{person}{Zhendong Wang}, \bibinfo{person}{Yaodi Liu}, \bibinfo{person}{Daojing He}, {and} \bibinfo{person}{Sammy Chan}.} \bibinfo{year}{2021}\natexlab{}.
\newblock \showarticletitle{Intrusion detection methods based on integrated deep learning model}.
\newblock \bibinfo{journal}{\emph{computers \& security}}  \bibinfo{volume}{103} (\bibinfo{year}{2021}), \bibinfo{pages}{102177}.
\newblock


\bibitem[Wang et~al\mbox{.}(2025)]%
        {wang2025cooperative}
\bibfield{author}{\bibinfo{person}{Zihan Wang}, \bibinfo{person}{Ziqi Zhao}, \bibinfo{person}{Yougang Lyu}, \bibinfo{person}{Zhumin Chen}, \bibinfo{person}{Maarten de Rijke}, {and} \bibinfo{person}{Zhaochun Ren}.} \bibinfo{year}{2025}\natexlab{}.
\newblock \showarticletitle{A cooperative multi-agent framework for zero-shot named entity recognition}. In \bibinfo{booktitle}{\emph{Proceedings of the ACM on Web Conference 2025}}. \bibinfo{pages}{4183--4195}.
\newblock


\bibitem[Wei et~al\mbox{.}(2022)]%
        {wei2022chain}
\bibfield{author}{\bibinfo{person}{Jason Wei}, \bibinfo{person}{Xuezhi Wang}, \bibinfo{person}{Dale Schuurmans}, \bibinfo{person}{Maarten Bosma}, \bibinfo{person}{Fei Xia}, \bibinfo{person}{Ed Chi}, \bibinfo{person}{Quoc~V Le}, \bibinfo{person}{Denny Zhou}, {et~al\mbox{.}}} \bibinfo{year}{2022}\natexlab{}.
\newblock \showarticletitle{Chain-of-thought prompting elicits reasoning in large language models}.
\newblock \bibinfo{journal}{\emph{Advances in neural information processing systems}}  \bibinfo{volume}{35} (\bibinfo{year}{2022}), \bibinfo{pages}{24824--24837}.
\newblock


\bibitem[Yang et~al\mbox{.}(2014)]%
        {yang2014combining}
\bibfield{author}{\bibinfo{person}{Lili Yang}, \bibinfo{person}{Jie Wang}, {and} \bibinfo{person}{Ping Zhong}.} \bibinfo{year}{2014}\natexlab{}.
\newblock \showarticletitle{Combining Supervised and Unsupervised Learning for automatic attack signature generation system}. In \bibinfo{booktitle}{\emph{International Conference on Algorithms and Architectures for Parallel Processing}}. Springer, \bibinfo{pages}{607--618}.
\newblock


\bibitem[Zhang et~al\mbox{.}(2022)]%
        {zhang2022adversarial}
\bibfield{author}{\bibinfo{person}{Chaoyun Zhang}, \bibinfo{person}{Xavier Costa-Perez}, {and} \bibinfo{person}{Paul Patras}.} \bibinfo{year}{2022}\natexlab{}.
\newblock \showarticletitle{Adversarial attacks against deep learning-based network intrusion detection systems and defense mechanisms}.
\newblock \bibinfo{journal}{\emph{IEEE/ACM Transactions on Networking}} \bibinfo{volume}{30}, \bibinfo{number}{3} (\bibinfo{year}{2022}), \bibinfo{pages}{1294--1311}.
\newblock


\bibitem[Zhang et~al\mbox{.}(2020)]%
        {zhang2020cmirgen}
\bibfield{author}{\bibinfo{person}{Runzi Zhang}, \bibinfo{person}{Mingkai Tong}, \bibinfo{person}{Lei Chen}, \bibinfo{person}{Jianxin Xue}, \bibinfo{person}{Wenmao Liu}, {and} \bibinfo{person}{Feng Xie}.} \bibinfo{year}{2020}\natexlab{}.
\newblock \showarticletitle{CMIRGen: Automatic signature generation algorithm for malicious network traffic}. In \bibinfo{booktitle}{\emph{2020 IEEE 19th International Conference on Trust, Security and Privacy in Computing and Communications (TrustCom)}}. IEEE, \bibinfo{pages}{736--743}.
\newblock


\end{thebibliography}

\appendix

\section{Discussion}
\textbf{System Architecture and Scalability.}
GRIDAI employs a multi-agent collaborative architecture comprising four LLM-based agents, respectively responsible for task coordination, rule generation, rule repair, and memory management. This modular design enhances the system’s interpretability and parallel processing capability, facilitating the automation of complex rule engineering workflows.
Nevertheless, such a multi-agent configuration inevitably introduces engineering challenges, including non-trivial communication overhead and scheduling complexity. When deployed under large-scale or real-time network traffic, these factors may affect operational efficiency and response latency. To address these potential bottlenecks, we are collaborating with an industrial partner to evaluate GRIDAI in production-like environments. Ongoing work focuses on enhancing scalability and runtime efficiency through distributed architectural optimization, caching and incremental inference to reduce redundant LLM calls, and hierarchical memory management to accelerate rule retrieval and update.

\textbf{Rational Scope of Experimental Design.}
The experiments in this study primarily aim to evaluate GRIDAI’s feasibility and effectiveness in automatic rule generation and repair. Since existing rule-generation methods generally lack repair functionality, the comparative experiments highlight GRIDAI’s design advantages and conceptual validity rather than providing a strictly parallel performance benchmark.
Due to data accessibility constraints, both the synthetic and real-world datasets used in the experiments remain moderate in size. Considering that attack samples in operational environments may arrive in arbitrary order, we conducted an additional evaluation using randomized sample sequences. As shown in Table \ref{tab:app3}, the randomized order had no significant impact on overall detection performance, although it slightly increased the Alert Duplication Rate (ADR). This suggests that GRIDAI’s decision process for sample–rule association is stable but could benefit from further refinement. Future research will explore incorporating retrieval-augmented generation (RAG) and rule semantic mapping techniques to enhance decision precision and robustness in dynamic traffic scenarios.

\begin{table}
   \caption{Performance of Randomly Input Samples}
   \label{tab:app3}
   \begin{tabular}{cccccc}
     \toprule
     \multirow{2}{*}{Order} & \multicolumn{5}{c}{Real-World Dataset}\\
     \cline{2-6}
      & Alerts $\downarrow$ & DA $\uparrow$ & DR $\uparrow$ & ADR $\uparrow$ & BAR $\downarrow$ \\
     \midrule
     ordered & 236 & 206 & 0.954 & 0.127 & 0.000\\
     random & 253 & 202 & 0.935 & 0.202 & 0.000\\
    
     \bottomrule
   \end{tabular}
\end{table}

\section{Dataset Details}
Due to data confidentiality restrictions, we are unable to release the dataset used in our experiments.
As an alternative, Table \ref{tab:app2} lists all Suricata rules (indexed by SID) employed for dataset filtering.
These rules are drawn from the latest Emerging Threats ruleset and cover a broad range of attack categories.

\begin{table}
  \centering
  \setlength{\tabcolsep}{6pt}
  \caption{Dataset Details.}
  \label{tab:app2}
  \begin{tabular}{c p{0.78\linewidth}}
    \toprule
    \textbf{Sid} & \multicolumn{1}{c}{\textbf{Attack Type}} \\
    \midrule
    \multicolumn{2}{c}{\textit{Synthetic Dataset}} \\
    \midrule
    2100993 & IIS admin access \\
    2016184 & ColdFusion admin access \\
    2101245 & ISAPI \texttt{.idq} exploit \\
    2009714 & XSS attempt (\texttt{</script>}) \\
    2009362 & Protected directory access (\texttt{/system32/}) \\
    2011696 & JBoss console WAR-upload exploit \\
    2101129 & \texttt{.htaccess} access (sensitive file) \\
    \midrule
    \multicolumn{2}{c}{{\textit{Real-World Dataset}}} \\
    \midrule
    2020093 & Neutrino malware communication (HTTP cookie) \\
    2009714 & XSS attempt (\texttt{</script>}) \\
    2024897 & Suspicious Go HTTP client User-Agent \\
    2017515 & Python requests User-Agent inbound \\
    2049400 & Sensitive file access (\texttt{/etc/passwd}) \\
    2027711 & Atlassian JIRA template injection RCE \\
    2006445 & SQL injection attempt (\texttt{SELECT FROM}) \\
    2030035 & AntSword webshell User-Agent \\
    2014017 & JBoss JMX console probe \\
    2027266 & Suspicious RAR file request \\
    2033403 & Apache SkyWalking GraphQL SQL injection \\
    2011768 & PHP code in HTTP POST (webshell upload) \\
    2025740 & Apache CouchDB RCE \\
    2010667 & Shell command execution attempt (\texttt{/bin/bash}) \\
    2016230 & WordPress plugin redirection vulnerability \\
    2024342 & Joomla SQL injection \\
    2034256 & Apache Shiro cookie deserialization RCE \\
    2033114 & Apache Solr DataImport handler RCE \\
    2024808 & Apache Tomcat JSP upload bypass \\
    2033452 & Kibana prototype pollution RCE \\
    2031190 & Nexus Repository Manager EL injection RCE \\
    2013031 & Suspicious Python-urllib User-Agent \\
    2030847 & Suspicious Windows Vista User-Agent \\
    2009715 & XSS attempt (\texttt{onmouseover}) \\
    2027251 & Suspicious DOC file request \\
    2027257 & Suspicious RTF file request \\
    2018056 & XXE attack via SYSTEM ENTITY \\
    2017260 & Generic ASP webshell upload \\
    2013508 & Downloader User-Agent activity \\
    2026916 & DirectsX malware C2 communication \\
    2101603 & HTTP DELETE method attempt (IIS) \\
    2013028 & curl User-Agent outbound \\
    2006447 & SQL injection attempt (\texttt{UPDATE SET}) \\
    2009485 & Sensitive file access (\texttt{/etc/shadow}) \\
    2027260 & Suspicious VBS file request \\
    2043026 & Suspicious empty Accept-Encoding header \\
    2010037 & SQL injection arbitrary file write (\texttt{INTO OUTFILE}) \\
    2027258 & Suspicious PS file request \\
    2017010 & SQL injection attempt (\texttt{xp\_cmdshell}) \\
    2018752 & Suspicious \texttt{.bin} file download from IP host \\
    2016415 & PHP injection in User-Agent \\
    2009375 & MSN chat activity (policy violation) \\
    2020912 & IIS Range header integer overflow DoS \\
    \bottomrule
  \end{tabular}
\end{table}

\section{Prompts}
We provide the complete prompt templates for the four agents in GRIDAI. 
Each template follows a standardized structure that defines the agent’s role and task instructions, ensuring transparent and reproducible multi-agent collaboration.

\begin{tcolorbox}[breakable, title={Prompt Template for \Aass}]
\texttt{\# YOUR ROLE}

- You are a cybersecurity expert specialized in analyzing and comparing network attack traffic. By examining behavioral and structural similarities, you can determine whether a new traffic payload represents a novel attack or a variant of an existing one.

- You are well-versed in common attack techniques, payload structures, evasion strategies, and signature-based intrusion detection.

- Your goal is to decide whether the new sample can be covered by modifying an existing rule or requires creating a new one.

\noindent\rule{\textwidth}{0.4pt}
\texttt{\# STAGE 1: RELATIONSHIP ASSESSMENT}

-   Below is the set of existing detection rules and their representative attack payloads, each rule corresponding to a distinct attack type or known variant: \textbf{<rule\_memory>}  
  
- The following is the new attack payload to be assessed: \textbf{<new\_payload>}

- Compare the new payload with all existing ones and decide whether it belongs to an existing attack family (variant) or constitutes a new attack.

- Follow these principles carefully:

1. Classify as a **variant** if the new payload shares the same attack method, exploited vulnerability, or payload structure with an existing one, and the corresponding rule could be slightly adjusted to detect it without affecting prior functionality.

2. Classify as a **new attack** if the new payload shares no consistent attack behavior with any existing sample and no existing rule can be modified to detect it without breaking other detections.

3. Focus on behavioral and structural similarities such as byte patterns and encoding or obfuscation methods.

4. Judge based on the underlying attack logic and execution process rather than textual similarity.

5. Mark a payload as a variant only when fully confident that an existing rule can be adapted to cover it.

- Your final output must strictly follow this format: **IsVariant, SID**,  
  where: IsVariant is True or False; SID is the corresponding rule ID if True, otherwise None.

\end{tcolorbox}

\begin{tcolorbox}[breakable, title={Prompt Template for \Agen}]
\texttt{\# YOUR ROLE}

- You are a network security expert specializing in analyzing malicious network traffic. You can extract anomalous indicators, attack behaviors, and exploited vulnerabilities from observed traffic. Based on this information, you can craft corresponding Suricata rules to protect the network from intrusion attempts.

- When asked to output the rules you’ve created, you need to provide a group of rules and save them in the ruleset file with the corresponding MCP tool. You must ensure that the rule group can pass the Suricata rule format check.

- Here is an example of Suricata rules:

** alert http any any -> any any (msg:"123"; flow: to\_server, established; http.uri; 
content:".abc"; nocase; classtype: attempted-recon; sid: 123;) **

\noindent\rule{\textwidth}{0.4pt}
\texttt{\# STAGE 1: PAYLOAD ANALYSIS}

- This is an attack traffic payload: \textbf{<payload>}.

- Please identify all possible anomalies, attack behaviors, and exploited vulnerabilities within it. Please pay extra attention to the following points:

1. Examine every potential vulnerability as comprehensively as possible—but avoid being overly broad.

2. You may cite open-source information (e.g., CVE, MITRE ATT\&CK). Highlight any reference used in your findings.

\noindent\rule{\textwidth}{0.4pt}
\texttt{\# STAGE 2: RULE GENERATION}

- Based on identified attack information, design corresponding Suricata rules. Each rule's ``msg'' field should include a brief rule description. Please pay extra attention to the following points:

1. Keep each rule concise and specific, targeting a single specific issue with sufficient applicability and robustness.

2. When using the “content” keyword to search, prefer short fields (e.g., "abc" rather than "def/abc") so the rule applies in more scenarios.

3. Handle request headers carefully and check them only when clearly related.

4. Avoid overly broad rules to reduce false positives and exclude meaningless patterns.

- After the generation is completed, save rules to the file \textbf{<ruleset.rules>}, and then call the MCP tool to verify whether rules can pass the format check and whether sample \textbf{<attack.pcap>} can trigger the correct alarms.

\noindent\rule{\textwidth}{0.4pt}
\texttt{\# STAGE 3: VALIDATION \& REGENERATION}

\texttt{Case 1 – Format Validation Failed:}

- The rules failed format check with message \textbf{<error>}.

- Regenerate and verify rules again.

\texttt{Case 2 – Rule Did Not Trigger Alerts:}

- The rules are correctly formatted but failed to detect the attack.

- Review the payload and regenerate effective detection rules as per previous requirements.

\noindent\rule{\textwidth}{0.4pt}
\# STAGE 4: OPTIMAL RULE SELECTION

- Among the rules just generated, the ones with SID \textbf{<sids>} triggered the correct alert.

- Please select the most optimal rule based on generalization, conciseness, robustness, and deployment value.
\end{tcolorbox}

\begin{tcolorbox}[breakable, title={Prompt Template for \Arep}]
\texttt{\# YOUR ROLE}

- You are a network security expert skilled in extracting abnormal indicators, attack behaviors, and exploited vulnerabilities from malicious network traffic and intrusion detection rules. You can also repair corresponding Suricata rules based on related information to protect the network from intrusion by attack variants.

\textit{(Rule output requirements are the same as those in \Agen.)}

\noindent\rule{\textwidth}{0.4pt}
\texttt{\# STAGE 1: PAYLOAD ANALYSIS}

- This is a set of attack traffic payloads: \textbf{<new\_payload; representative\_attack\_payload>}.

- They correspond to the same type of attack behavior and its variants. Please identify all possible anomalies, attack behaviors, and exploited vulnerabilities within it. Please pay extra attention to the following points:

1. Focus on the common issues present across this set of payloads, as they correspond to the same attack technique.

\textit{(Remaining requirements are the same as those in \Agen.)}

\noindent\rule{\textwidth}{0.4pt}
\texttt{\# STAGE 2: RULE REPAIR}

- Based on identified information, repair the corresponding Suricata rule \textbf{<old\_rule>}. It is partially effective but fails to match all payloads. Each rule's ``msg'' field should include a brief description. Please pay extra attention to the following points:

1. Focus on identifying shared anomalies or patterns that consistently appear across all payloads.

\textit{(Remaining requirements are the same as those in \Agen.)}

- After the repair is completed, save rules to the specified file \textbf{<ruleset.rules>}, and then call the MCP tool to verify whether rules can pass the format check and whether the samples \textbf{<attack.pcap; representative\_attack.pcap>} can all trigger the correct alarms.

\noindent\rule{\textwidth}{0.4pt}
\texttt{\# STAGE 3: VALIDATION \& REGENERATION}

\texttt{Case 1 – Format Validation Failed:}

\textit{(Remaining requirements are the same as those in \Agen.)}

\texttt{Case 2 – Rule Did Not Trigger Alerts:}

- The rules are correctly formatted but failed to trigger alerts for all attack samples.

- For \textbf{<success.pcap>}, Suricata output an alarm as expected. For \textbf{<fail.pcap>}, updated rules are ineffective.

- Review the payloads and regenerate effective detection rules as per previous requirements.

\noindent\rule{\textwidth}{0.4pt}
\# STAGE 4: OPTIMAL RULE SELECTION

\textit{(Remaining requirements are the same as those in \Agen.)}

\end{tcolorbox}

\begin{tcolorbox}[breakable, title={Prompt Template for \Amem}]
\texttt{\# YOUR ROLE}

- You are a cybersecurity expert skilled in analyzing malicious traffic and Suricata detection rules to understand their underlying attack behaviors, structures, and patterns.

- Your goal is to identify, among multiple samples detected by the same rule, the most representative and valuable attack sample to serve as that rule’s reference in the rule memory repository.

\noindent\rule{\textwidth}{0.4pt}
\texttt{STAGE 1: SAMPLE SELECTION}

- Below is a Suricata rule and several attack payloads it successfully detects:

\textbf{<new\_rule; payloads>}

- All samples have been verified to trigger correct alerts and can serve as valid references.

- From the perspective of a professional security analyst, determine which sample best represents the attack behavior covered by this rule. Evaluate each sample based on representativeness, clarity, coverage, analytical value.

- Select one sample that best meets these criteria. After selection, invoke the MCP tool to update the corresponding RuleItem in the rule memory repository.

\end{tcolorbox}

\end{document}